\documentclass[11pt,oneside]{article}

\usepackage{natbib}
\bibpunct[, ]{(}{)}{,}{a}{}{,}%
\usepackage{graphicx}
\usepackage{amsmath}
\usepackage{amssymb}
\usepackage{amsthm}
\usepackage{tikz}
\usetikzlibrary{shapes}
\usetikzlibrary{patterns}
\usetikzlibrary{arrows,positioning,decorations.pathmorphing,trees}
\usetikzlibrary{arrows.meta}
\usepackage{tikz-cd}
\usepackage{mathtools}
\usepackage{stackengine}
\usepackage{pgfplots}
\usepackage{float}
\usepackage{braket}
\usepackage{eurosym}
\usepackage{hyperref}
\usepackage{comment}
\usepackage{subcaption}
\usepackage{ifthen}

\newcommand{\ch}{\overline{conv}}
\newcommand{\conch}{\overline{conc}}

\usepackage{fullpage}

\newtheorem{theorem}{Theorem}[section]

\newtheorem{corollary}[theorem]{Corollary}

\newtheorem{proposition}[theorem]{Proposition}
\newtheorem{definition}{Definition}

\newtheorem{example}{Example}
\newboolean{includeHidden}
\setboolean{includeHidden}{false}
\newcommand{\extR}{\overline{\mathbb{R}}}
\newcommand{\dom}{\textnormal{dom}}

\newboolean{tablesInText}
\setboolean{tablesInText}{true}
\newboolean{figuresInText}
\setboolean{figuresInText}{true}
\newboolean{footnotesInBrackets}
\setboolean{footnotesInBrackets}{true}

\newtheorem{result}{Result}
\usepackage{lipsum}

\def\argmax{\mathop{\rm arg\,max}}%

\title{Pricing Optimal Outcomes in Coupled and Non-Convex Markets: Theory and Applications to Electricity Markets}
\author{Mete \c{S}eref Ahunbay \and Martin Bichler \and Johannes Knörr\thanks{Technical University of Munich, School of Computation, Information and Technology, Department of Computer Science. E-mail: \texttt{mete.ahunbay/bichler/knoerr@cit.tum.de}}}

\begin{document}
	
\maketitle

\begin{abstract}
	According to the fundamental theorems of welfare economics, any competitive equilibrium is Pareto efficient.
	Unfortunately, competitive equilibrium prices only exist under strong assumptions such as perfectly divisible goods and convex preferences. 
	In many real-world markets, participants have non-convex preferences and the allocation problem needs to consider complex constraints. 
	Electricity markets are a prime example, but similar problems appear in many real-world markets, which has led to a growing literature in market design. Power markets use heuristic pricing rules based on the dual of a relaxed allocation problem today. 
	With increasing levels of renewables, these rules have come under scrutiny as they lead to high out-of-market side-payments to 
	some participants and to inadequate congestion signals. 
	We show that existing pricing heuristics optimize specific design goals that can be conflicting. 
	The trade-offs can be substantial, and we establish that the design of pricing rules is fundamentally a multi-objective 
	optimization problem addressing different incentives. 
	In addition to traditional multi-objective optimization techniques using weighing of individual objectives, 
	we introduce a novel parameter-free pricing rule that minimizes incentives for market participants to deviate locally.
	Our theoretical and experimental findings show how the new pricing rule capitalizes on the upsides of existing pricing rules under scrutiny today. It leads to prices that incur low make-whole payments while providing adequate congestion signals and low lost opportunity costs. Our suggested pricing rule does not require weighing of objectives, it is computationally scalable, and balances trade-offs in a principled manner, addressing an important policy issue in electricity markets. 
 \end{abstract}

\section{Introduction} \label{sec:Intro}

Low transaction costs on electronic markets have led to an increased use of market mechanisms to allocate scarce resources. The promise of markets rests on fundamental theoretical results such as the welfare theorems \citep{arrow1954existence}. 
These theorems state that in markets with convex preferences, a Walrasian equilibrium will maximize social welfare and that every welfare-maximizing allocation can be supported by Walrasian equilibrium prices. At those prices no market participant would want to deviate from what they are assigned, and the outcome is envy-free and budget-balanced. 
The welfare theorems provide the theoretical rationale for using markets in a wide variety of applications ranging from the sale of frequency spectrum to treasury bills. 

However, more recent literature on competitive equilibrium theory for markets with multiple indivisible goods shows that Walrasian equilibrium prices only exist under restrictive assumptions on the valuation functions \citep{bikhchandani1997competitive, Gul99, bikhchandani2002package, baldwin2019understanding}. Most real-world markets have non-convex value and cost functions leading to non-convex welfare maximization problems, where Walrasian equilibrium prices do not exist in general. Electricity markets are a prime example of such non-convex markets, where ramp-up costs for generators or block bids of buyers lead to significant non-convexities. 
Similar characteristics can be found in industrial procurement, spectrum auctions, and transportation markets \citep{cramton2006combinatorial}. As a result, pricing in non-convex markets has become a central topic in market design \citep{Milgrom.2017}.
 
While the fundamental problems that arise in pricing goods in non-convex markets are independent of the domain, electricity market design is particularly challenging due to the fact that supply and demand need to be balanced at all times and due to specific bid languages. These bidding languages account for operational restrictions and complex preference functions of market participants and thereby imply non-convexities \citep{Herrero.2015, Kuang.2019}. More importantly, electricity markets consist of coupled markets where trade happens on interlinked nodes in the electricity grid. As a result, congestion on network links needs to be reflected in the prices as well. These specifics lead to one of the most challenging market design problems, one that is receiving renewed attention due to the energy transition that is currently ongoing world-wide \citep{tian2022global}. 
Currently, market operators in U.S. electricity markets determine the welfare-maximizing dispatch by solving a non-convex allocation problem, including network constraints, and obtain nodal prices through convexification of the original problem (e.g., Convex Hull or IP pricing, cf. Section \ref{sec:Theory}). The outcome does not constitute a Walrasian equilibrium and out-of-market side-payments are required to compensate for incentives to deviate from the dispatch. 
In fact, current pricing rules have come under scrutiny since these side-payments to some of the market participants have become very high. The U.S. FERC recently argued that ``the use of side-payments can undermine the market’s ability to send actionable price signals.''\footnote{\url{https://www.ferc.gov/industries-data/electric/electric-power-markets/energy-price-formation}} The resulting price signals are considered intransparent, and they can set flawed incentives as they do not reflect marginal costs anymore. 

\subsection{Multiple Pricing Objectives}

We consider the following auction mechanism: a market operator collects bids and asks from market participants and solves for the optimal, i.e. welfare-maximizing, allocation of different items (\textit{the allocation problem}). Subsequently, the market operator determines prices of items (\textit{the pricing problem}) and makes allocation and prices public. In a Walrasian equilibrium it is possible to determine linear and anonymous prices that align all incentives such that no participant wants to deviate from their outcome (aka. envy-freeness) and the market is budget balanced. 
Unfortunately, such equilibria generally only exist in convex markets \citep{starr1969quasi, bikhchandani1997competitive}. 
Therefore, in markets with non-convexities, one must resort to other pricing rules that violate either envy-freeness or budget-balance.
We refer to incentives to deviate from the optimal outcome as \textit{global lost opportunity costs}, GLOCs. They describe the difference between each participant's profits under the welfare-maximizing allocation and the individual profit maxima each participant could obtain given the prices. Most prior work on pricing in non-convex electricity markets has focused on minimizing GLOCs and thereby approximating a Walrasian equilibrium. This pricing rule is referred to as \emph{Convex Hull} (CH) \emph{pricing} \citep{gribik2007market, hogan2003minimum}. The market operator then implements the optimal outcome and either compensates participants for some or all of their GLOCs (violating budget-balance) or accepts that participants have incentives to deviate (violating envy-freeness). 

This paper is based on the observation that in non-convex electricity markets a market operator needs to consider multiple incentives to deviate from the optimal allocation. In particular, while GLOCs capture all incentives to deviate from the optimal allocation, market participants are typically not compensated for their \textit{entire} GLOCs. Instead, additional compensations in the form of individual side-payments are only paid out to those market participants that incur a loss under the given prices, addressing individual rationality.\footnote{The remaining GLOCs are typically enforced by imposing penalties on market participants that deviate from the optimal allocation.} These side-payments are called \textit{make-whole payments}, MWPs. As MWPs have become very high, they have been a concern to market operators and regulators alike. 
Developing pricing rules that reduce MWPs has been the focus of much recent research \citep{Bichler.2021, ONeill.2019}. We refer to this pricing rule as \emph{minimum MWP} (Min-MWP) \emph{pricing}. MWPs are a subset of GLOCs, but neither does Convex Hull pricing always lead to small MWPs nor does minimum MWP pricing lead to small GLOCs, as we will show. 

A specific aspect of electricity markets is that they are based on a network with possibly thousands of spatially distributed nodes connected via transmission lines, operated by transmission system operators. To put it differently, we have a set of coupled markets where supply and demand are equalized by participants who only provide transmission. Thus, we also need to price transmission lines in such coupled markets appropriately \citep{Lete.2022}. For example, prices should only differ across a pair of nodes if there is \emph{congestion} in the network and no further power may be transmitted from the node with the higher price to that with a lower price. Violations of this condition, i.e. price differences across uncongested branches, may result in a product revenue shortfall for transmission operators which would require compensation. Unfortunately, neither Convex Hull pricing nor minimum MWP pricing satisfy this requirement \citep{Schiro.2016, Bichler.2021}. We therefore introduce a third class of incentives to deviate from the optimal allocation. \textit{Local lost opportunity costs}, LLOCs, measure incentives to deviate from the optimal allocation \textit{under fixed commitment}.\footnote{Commitment decisions on spot markets determine whether a generator is scheduled to produce electricity during a market time unit (a binary decision variable in the allocation problem), but not the production quantity (in Megawatt hours (MWh)). Commitment decisions occur because many types of generators require a long time to turn on/off.} In other words, LLOCs assume that the commitment decisions have been made but that generators can deviate from their assigned volumes in an attempt to improve their payoff.
Minimizing LLOCs also provides for the desired congestion signals in the network, in the sense that prices reflect the marginal value of additional transmission capacity \citep{Yang.2019}. 
A pricing rule that yields zero LLOCs is the well-known \emph{integer programming} (IP) \emph{pricing} \citep{oneill2005efficient}. Being widely used in U.S. electricity markets, this pricing rules involves solving for the optimal allocation, fixing binary commitment variables to their optimal values, and solving the dual of the resulting linear program to obtain prices.

\subsection{Contributions}

It is well known that in a convex market, one can derive Walrasian equilibrium prices from the dual variables of the welfare maximization problem and these prices have zero GLOCs, LLOCs and MWPs. First, we generalize the welfare theorems and show that they hold in coupled markets as long as the preferences of buyers and sellers are convex. We prove the theorem for any convex cost or concave valuation functions, which do not need to be differentiable. This delineates environments where we can expect to find a Walrasian equilibrium from those (non-convex) markets where this is not the case. 

Second, we focus on non-convexities in electricity spot markets and reconsider the pricing problem based on the specifics of these coupled and non-convex markets. Minimizing envy (in the form of GLOCs) has been the guiding principle of Convex Hull pricing, which is widely considered as central approach to electricity market pricing. 
However, pure minimization of GLOCs as in Convex Hull pricing is computationally intractable and can lead to unreasonably high LLOCs and MWPs. This causes high side-payments and budget-balance is often violated. More importantly, wrong congestion signals are a major concern as they can trigger demand response in cases where this is not necessary or vice versa. Congestion signals refer to the ability of prices to reflect congestion on the transmission lines between the nodes. More accurately, prices between nodes should only differ when congestion occurs in the network and the price difference reflects the marginal value of transmission capacity. 

More generally, low GLOCs, LLOCs, and MWPs are conflicting design goals and optimizing only one of these classes comes at the expense of another. 
If MWPs are minimized, GLOCs can be unreasonably high and congestion signals become distorted leading to high LLOCs. Focusing only on LLOCs can be equally harmful and lead to very high MWPs for some participants, a phenomenon that has caught the interest of policy makers. In a second contribution, we establish these trade-offs for market operators and formalize the pricing problem as a multi-objective optimization problem. This breaks with prior literature that mainly focuses on GLOCs in electricity market pricing. By using our convex optimization framework, we also make explicit the link between previously proposed pricing rules and the classes of incentives they optimize. 

In our third contribution we introduce the join pricing rule, which balances trade-offs between MWPs and LLOCs to minimize local incentives of participants to deviate from the efficient outcome as well as incentives to exit the market. As discussed earlier, the first objective addresses adequate congestion signals, while the second objective avoids high side-payments by the market operator.  
We prove that the join pricing rule always achieves lower MWPs than IP pricing, and lower LLOCs than any minimum MWP solution if zero MWPs are attainable. In addition, we show that prices computed via the join result in a participant-wise Pareto-optimal outcome, such that different prices cannot jointly reduce MWPs and LLOCs of this participant. 
As a practical advantage, the join does not require to specify weights and is thus a parameter-free pricing rule. The join is different to other techniques that have been discussed in the extensive literature on multi-objective optimization, and contributes to this literature as well \citep{jahn2009vector}.

In addition to the formal characterization, we show in extensive numerical experiments that prices computed via this join require significantly less MWPs than traditional IP pricing and retain good congestion signals with very low LLOCs at the same time. Besides, the approach can be computed efficiently and it requires no fine-tuning of objective weights, as pointed out earlier. In electricity markets, where global incentives to deviate (GLOCs) can be enforced by penalizing deviations, the join strikes a desirable balance, and the experiments show that the remaining incentives to deviate are low.
Our findings have implications for the ongoing electricity pricing discussions in the U.S. and in the EU, and we make a novel proposal that promises significant advantages regarding MWPs and the quality of congestion signals. 

\subsection{Organization of the Paper}
The remainder of this paper is structured as follows. Section \ref{sec:lit} summarizes relevant literature on competitive equilibrium theory and on electricity market design. 
In Section \ref{sec:PricingCoupled}, we introduce a generic market and revise central findings for convex and non-convex settings. In Section \ref{sec:Theory}, we outline how current pricing rules each optimize different objectives, and how these objectives can be in substantial conflict with each other. To that end, we propose a multi-optimization perspective in Section \ref{sec:combinations} and introduce principled ways to balance the trade-offs. Subsequently, we tailor these pricing rules to an exemplary electricity market in Section \ref{sec:Electricity}. We present explicit formulations and numerical findings that illustrate the advantages of our pricing rules. Section \ref{sec:Conclusion} provides a summary and conclusions. 

\section{Related Literature}\label{sec:lit}

The literature on competitive equilibrium has a long history. In this section, we summarize the main theoretical findings before we discuss the related literature on electricity market design. These streams of literature are often considered separately, while we aim to connect the contributions of economics and engineering.

\subsection{Competitive Equilibrium Theory}

Early in the study of markets, general equilibrium theory was used to understand how markets could be explained through the demand, supply, and prices of multiple commodities. The Arrow-Debreu model shows that under convex preferences, perfect competition, and demand independence there must be a set of competitive equilibrium prices \citep{arrow1954existence}. Market participants are price-takers, and they sell or buy goods in order to maximize their total utility. 
General equilibrium theory assumes divisible goods and convex preferences and the well-known welfare theorems do not carry over to markets with indivisible goods and complex (non-convex) preferences and constraints. 

More recently, competitive equilibria with indivisible objects were studied and the idea of a quasilinear utility function was widely adopted \citep{baldwin2019understanding, bikhchandani1997competitive, bikhchandani2002package}. In these competitive equilibrium models, buyers and sellers with a quasilinear utility maximize their respective payoffs at the prices, resulting in an outcome that is stable (i.e. no participant wants to deviate from their resulting trade). A large part of the literature focuses on Walrasian equilibria, i.e. efficient market outcomes with linear (item-level) and anonymous prices, where all participants maximize payoff. If such prices exist, then the outcome is allocatively efficient, i.e. it maximizes welfare, as can be shown via linear programming duality. 
In general, Walrasian equilibria for markets with indivisible goods only exist for restricted valuations such as strong substitutes \citep{baldwin2019understanding, bikhchandani1997competitive, Kelso82}. These conditions lead to a concave aggregate value function, the allocation problem can be solved in polynomial time, and linear and anonymous (Walrasian) competitive equilibrium prices clear the market \citep{bichler2019computing}. Importantly, under these conditions the welfare theorems hold for markets with indivisible objects \citep{blumrosen2007combinatorial, bichler2020walrasian}. 

Unfortunately, these conditions are very restrictive and in most markets goods can be substitutes and complements such that no Walrasian equilibria exist. Besides, most allocation problems that have been analyzed in operations research and the management sciences (e.g., various types of scheduling or packing problems) lead to NP-hard combinatorial optimization problems, which can neither be solved in polynomial time nor do they exhibit simple Walrasian equilibrium prices. This has led to significant research on non-convex combinatorial markets, which allow bidders to specify package bids, i.e. a price is defined for a subset of the items \citep{bichler2017handbook, Milgrom.2017}. The specified bid price is only valid for the entire package and the package is indivisible such that bidders can express complex (quasilinear) preferences for general valuations including complements and substitutes. 

The generality of these markets comes at a price. First, the winner determination problem becomes an NP-hard optimization problem. Second, competitive equilibrium prices need to be non-linear and personalized to allow for full efficiency \citep{bikhchandani2002package}. \citet{bichler2017core} show that the core of the game can even be empty such that no competitive equilibrium prices exist. 
\citet{Milgrom.2022} recently introduced near-efficient mechanisms for non-convex markets that are computationally scalable, nearly incentive-compatible, and produce two price vectors (one for sellers and one for buyers). 
However, in practice, non-linear and personalized prices would convey little information other than that a bidder lost or won. Besides, if prices should serve as a baseline for futures trading, this is hardly possible with non-linear prices that differ among participants. In other words, anonymity and linearity are important requirements for many markets, electricity markets being the prime example. 

\subsection{Pricing on Electricity Markets}

Electricity markets typically comprise a sequence of markets. The characteristic network structure of these markets has a major influence on market design decisions \citep{Ehrenmann.2009}. While forward \citep{Dent.2011, Peura.2021} and reserve markets \citep{Papavasiliou.2011} are highly important, we focus on the central wholesale spot markets that are based on auctions. Here, market participants submit supply and demand bids according to a certain \textit{bid language} which translates into a central allocation problem. As market participants often exhibit start-up costs, minimum generation requirements, or other technical constraints, bid languages typically imply some form of non-convexities \citep{herrero2020evolving}. With the advent of variable energy sources, demand response becomes increasingly important \citep{Papavasiliou.2014}. To adequately reflect flexibility on the demand side, market operators need new bid formats that likely lead to additional non-convexities and price-sensitive demand \citep{Bichler.2021}. Further, the multi-period nature of the clearing problem adds complexity \citep{Cho.2022}. Market operators around the world use mixed-integer programming (MIP) to address these non-convexities and to determine the efficient allocation or dispatch \citep{Hillier.2001}. Even stochastic versions of such models have been proposed \citep{Papavasiliou.2013, Zavala.2017}. However, computing (electricity) prices in the presence of non-convexities remains a fundamental problem. 

If Walrasian equilibrium prices exist, no market participant will have an incentive to deviate from the optimal allocation. In other words, no participant would bear GLOCs. A natural extension to non-convex markets -- where Walrasian equilibrium prices do not exist in general -- is to minimize these GLOCs but maintain linearity and anonymity of prices. This is referred to as \emph{Convex Hull (CH) pricing}, originally explored by \citet{gribik2007market} based on \citet{hogan2003minimum}. Convex Hull pricing replaces the non-convex feasible region of the combinatorial allocation problem by its convex hull, and obtains prices from the dual of the resulting convex problem. We refer to \citet{Schiro.2016} for a comprehensive and critical overview. 

However, obtaining exact Convex Hull prices is computationally expensive. A common approach involves solving the (convex but non-smooth) Lagrangian dual of the original non-convex problem, which -- under mild assumptions -- is equivalent to optimizing the convex envelope of the original cost functions over the convex hull of the feasible region \citep{falk1969lagrange, Hua.2017}. Several algorithms have been introduced, including subgradient methods \citep{Ito.2013}, extreme-point subdifferential methods \citep{Wang.2013}, and many more \citep{Wang.2013.Subgradient, Goffin.2002, Yu.2020, Knueven.2022, stevens2022application}. Convex Hull pricing has also been shown to be robust when exact solutions to the (NP-hard) welfare maximization problem cannot be computed \citep{Eldridge.2020}. Overall, there has been significant progress in this field, but the methods are not yet used in practice and computational complexity remains a major concern. 

In practice, market operators resort to different heuristics in order to price electricity on real-world markets. Most practical implementations are based on \emph{integer programming (IP) pricing}, where the non-convex allocation problem is first solved to optimality, and then solved again with integer variables being fixed to their optimal values. The IP prices are derived from the dual solution of the latter convex program \citep{oneill2005efficient}. IP pricing has become popular as it follows the notion of marginal cost pricing in non-convex markets and furthermore provides accurate congestion signals when applied on an electricity network. However, IP prices do not constitute competitive equilibrium prices and can come with high lost opportunity costs. In an attempt to address this, some U.S. markets have started to apply \emph{Extended Locational Marginal pricing} (ELMP), where prices are based on the dual variables of the linear programming relaxation of the non-convex allocation problem \citep{MISO.2019}. The motivation to use ELMP is based on the finding that under certain assumptions on the bidding formats, ELMP corresponds to the primal approach to compute Convex Hull prices \citep{Hua.2017}. However, this property generally does not hold, and Convex Hull prices as well as ELMP can imply high MWPs. As a result, new pricing rules have been proposed in an attempt to reduce MWPs \citep{ONeill.2019, Bichler.2021}, either via price differentiation or by minimizing them directly. Recently, \citet{Yang.2019} proposed to consider congestion signals in the design of pricing rules, which we do in our proposal. A comprehensive discussion of pricing rules as they have been proposed in the academic literature is beyond this paper. We point the interested reader to \citet{liberopoulos2016critical}, who provide an excellent overview of the relevant literature.

\section{Preliminaries} \label{sec:PricingCoupled}

In this section, we introduce a model for a non-smooth and coupled market which allows for Walrasian equilibria. 
Much of the prior literature on pricing in non-convex markets relies on linear programming (LP) and LP duality \citep{bikhchandani1997competitive, liberopoulos2016critical}. 
Our paper draws in large parts on the more general field of convex optimization allowing for general convex and not just linear cost or value function. Therefore, we first provide a brief introduction on the relevant notions from convex and non-smooth optimization \citep{bagirov2014introduction}.

\subsection{Basics of Convex Optimization} \label{sec:Prelim}

We work with functions that take values in the extended real line, $\extR = \mathbb{R} \cup \{-\infty,\infty\}$. Addition and multiplication are commutative binary operations on the extended real line, defined as usual for any real number, and $x + \infty  = \infty$, $x - \infty  = -\infty$ for any $x \in \mathbb{R}$.

\begin{definition}[Convexity, closedness and properness]
A function $f : \mathbb{R}^n \rightarrow \extR$ is called
\begin{enumerate}
\item \textbf{convex} if for any $x, y \in \mathbb{R}^n$, for any $\lambda \in [0,1]$, $f(\lambda x + (1-\lambda) y ) \leq \lambda f(x) + (1-\lambda) f(y)$,
\item \textbf{closed} (or \textbf{lower semi-continuous}) if for any $x \in \mathbb{R}^n$, $\liminf_{y \rightarrow x} f(y) \geq f(x)$, and
\item \textbf{proper} if $f(x) \neq -\infty$ for any $x \in \mathbb{R}^n$, and $f \neq \infty$ identically.
\end{enumerate}
In turn, a set $S \subseteq \mathbb{R}^n$ is called convex if its characteristic function $\chi_S$ is convex, where $\chi_S(x) = 1$ if $x \in S$ and $0$ otherwise. 
\end{definition} 

A function $f$ might not be closed or convex, but it is always possible to consider its convex closure. The \textbf{convex closure} of a function $f$, $\ch(f)$, is the pointwise maximum closed and convex function that underestimates it,
$$ \ch(f)(x) = \max \{ g(x) | g \leq f \text{ is closed and convex} \}.$$

A closed, proper and convex function $f$ admits information on the change in the value of the function in response to small changes at each point in its domain. This quantity is given in the form of an \emph{affine minorant}, and is the analogue of a gradient for differentiable functions.

\begin{definition}[Affine minorants and subgradients]\label{def:subgradient}
A function $f : \mathbb{R}^n \rightarrow \extR$ is said to have an \textbf{affine minorant} if $\exists p \in \mathbb{R}^n, c \in \mathbb{R}, \forall x, f(x) \geq p^T x + c$. If $c = f(x^*) - p^T x^*$ for some $x^* \in \mathbb{R}^n$, i.e. if 
$$ f(x) \geq f(x^*) + p^T (x - x^*) \ \forall \ x \in \mathbb{R}^n,$$
then $p$ is said to be a \textbf{subgradient} of $f$ at $x^*$. The set of all subgradients of $f$ at a given point $x$ is called the \textbf{subdifferential} of $f$ at $x$, and is denoted $\partial f(x)$. 
\end{definition}

For a function $f$ that has an affine minorant, a transformation of $f$ encodes information on the set of its affine minorants. This transformation is obtained by evaluation of a maximization problem.

\begin{definition}\label{def:conjugate}
For a function $f : \mathbb{R}^n \rightarrow \extR$, the \textbf{Legendre-Fenchel transformation} (or the \textbf{convex conjugate}) of $f$ is the function
$$ f^*(p) = \sup_{x \in \mathbb{R}^n} p^T x - f(x).$$
\end{definition}

In particular, for a proper function $f$, $f^*(p) < \infty$ if and only if there exists $c$ such that $p^T x + c$ is an affine minorant of $f$. In this case, $f^*$ is closed, convex and proper. Furthermore, the biconjugate $f^{**}$ then equals $\ch(f)$, the convex closure of $f$ \citep{rockafellar2015convex}.
Convex conjugation provides a connection between operations on functions $f, g : \mathbb{R}^n \rightarrow \extR$ and their conjugates $f^*, g^*$. We remark that for a function $f$ taking values in a \emph{primal} space, its conjugate $f^*$ has its arguments in the corresponding \emph{dual} space. For instance, in our economic setting, as we shall see in (\ref{def:indirect-utility}), if $f(x)$ corresponds to the costs a generator incurs for supplying $x$ MWh then $f^*(p)$ is the maximum profit attainable by this generator given a price vector $p$  -- i.e. the generator's \emph{indirect utility}.

The following propositions are standard in convex analysis, and we need them to prove the welfare theorems for coupled markets and in the analysis of the pricing rules we consider in this paper. The first proposition lists the rules for the algebraic manipulation of functions and their conjugates, which we employ to study the link between pricing rules and their associated convex market models.

\begin{proposition}[Calculus of convex conjugation \citep{rockafellar2015convex}]\label{prop:calculus}
	Let $f, g: \mathbb{R}^n \rightarrow \extR$, $\alpha, \beta \in \mathbb{R}$ such that $\alpha > 0$, and $v \in \mathbb{R}^n$, then the following hold:
	\begin{enumerate}
		\item (Translation by a vector) Addition of a linear function to $f$ corresponds to a constant shift in the argument of its conjugate, $(f - v^T (\cdot) )^* = f^*((\cdot)+v)$.
		\item (Addition of a number) Addition of a constant to $f$ corresponds to subtracting the same constant from its conjugate, $(f + \beta)^* = f^* - \beta$.
		\item (Multiplication by a number) Rescaling $f$ by some positive constant rescales both the magnitude and the argument of its conjugate, $(\alpha f)^* = \alpha f \left( \frac{(\cdot)}{\alpha} \right)$.
		\item (Convolution) Suppose the function $\omega : \mathbb{R}^n \rightarrow \extR$ is defined via a convolution of $f$ and $g$, i.e.  
		$$\omega(\sigma) = \min_{Ax + By = \sigma} f (x) + g(y)$$
		for some $m \times n$ matrices $A,B$. Then for any $p \in \mathbb{R}^m$ the conjugate of $\omega$ is given by 
		$$\omega^*(p) = f^*(A^T p ) + g^* (B^T p).$$ 
		\item (Partial closure) The convex closure of the minimum of two functions $f,g$ has as its conjugate the pointwise maximum of the conjugates $f^*, g^*$, $(\ch \min \{f, g\})^* = \max\{f^*, g^*\}$.
	\end{enumerate}
\end{proposition}
Specifically, in the next section we will see that optimal market clearing subject to supply-demand equivalence is a maximization problem, which can be interpreted as the value of a welfare function defined by a convolution. Thus Proposition \ref{prop:calculus}.4 implies that the conjugate of this welfare function additively separates over participants' indirect utility functions. Pricing problems in practice turn out to be approximations of this conjugate, and the rest of Proposition \ref{prop:calculus} allows us study their corresponding convex models. 

The next proposition shows that if $f$ is closed and convex, then its convex conjugate encodes information on its subgradients:

\begin{proposition}[Fenchel-Young (in)equality]
Suppose that $f$ is closed, convex and proper. Then for any $p,x \in \mathbb{R}^n$,
$$f^*(p) + f(x) \geq p^T x.$$
Moreover, the inequality holds with equality if and only if $p \in \partial f(x)$.
\end{proposition}

Thus for any $x$, $\partial f(x) = \argmax_p p^T x - f^*(p)$. In particular, if $f(x)$ is given as the solution to a linear (minimization) program, then the subdifferential $\partial f(x)$ is a polyhedron whose form may be noted down explicitly.

We conclude this section by noting that the results above may be extended immediately to concave functions. A function $f$ is called \text{concave} if $-f$ is convex, and \textbf{upper semi-continuous} if $-f$ is closed. The \textbf{concave closure} $\conch(f)$ is then simply $-\ch(-f)$, and the definitions of \emph{affine majorants} and \emph{supragradients} follow analogously to the discussion above.

\subsection{Dual Pricing in Convex Markets} \label{sec:Convex_Coupled_Markets}

Our goal in this paper is to gain an understanding of previously proposed pricing rules through the lens of convex optimization to reveal the corresponding design objectives they optimize. Towards this end, in this section, we first introduce our market model and then formalize the notion of a dual pricing problem via analysis of a convex market, where there exists a canonical pricing rule for optimal outcomes -- Walrasian equilibria.

\begin{definition}
A \textbf{coupled market} consists of a set of goods $M$, a set of transmission network parameters $F$ and a set of market participants $L = B \cup S \cup R$, partitioned into the set of buyers $B$, set of sellers $S$ and the set of transmission operators $R$. Each market participant $\ell \in L$ has preferences over bundles in $\mathbb{R}^{M \cup F}$, i.e. each buyer $b$ has a valuation function $v_b : \mathbb{R}^{M \cup F} \rightarrow \extR$, each seller $s$ has a cost function $c_s : \mathbb{R}^{M \cup F} \rightarrow \extR$, and each transmission operator $r$ has a cost function $d_r : \mathbb{R}^{M \cup F} \rightarrow \extR$. 
\end{definition}

The set of goods $M$ may also include identical products at different points in time, e.g., electricity at the same node but for different hours. Transmission operators perform trades between different nodes in the network and are modeled as a separate class of market participants. We encode any feasibility constraint for market participants in the domain of cost functions, i.e. buyers have valuation $-\infty$ and sellers / transmission operators have cost $+\infty$ for an infeasible bundle. For now we assume concave valuation and convex cost functions (e.g., sellers incur no fixed costs when selling a positive amount). We typically write $x_b \geq 0$ for a bundle purchased by buyer $b$, $y_s \geq 0$ for a bundle supplied by seller $s$ and $f_r \in \mathbb{R}$ for an exchange enacted by transmission operator $r$ (the sign of $f_r$ indicates the direction of the exchange). To make the notation more concise, we write $z_\ell$ for the allocation of an arbitrary market participant $\ell \in L$. Finally, we assume that buyers and sellers only have values for the consumption or supply of goods in $M$ while interactions over the transmission network are performed exclusively by transmission operators, i.e. only bundles in $\mathbb{R}^{M} \times \{0\}^F$ are feasible for buyers and sellers.

As a result of encoding feasibility in costs, allocations to market participants are assumed to be constrained only by the set of supply-demand equivalence constraints
\begin{equation}\label{cons:supply-demand-eq}
\sum_{s \in S} y_s - \sum_{b \in B} x_b + \sum_{r \in R} B_r f_r = 0,
\end{equation}
where $B_r$ is some matrix specifying how flows in the transmission network interact with the supply-demand balance for the goods at each market. For example, for the DCOPF model considered below, this matrix indicates how current that is injected at a node will distribute over the transmission lines, and is determined by the line susceptances. 

We are concerned with supporting an \emph{optimal allocation} with \emph{prices}. An optimal allocation is a solution $(z^*_\ell)_{\ell \in L}$ of the \emph{welfare maximization problem}
\begin{equation}\max_{x,y,f}  \sum_{b \in B} v_b(x_b) - \sum_{s \in S} c_s(y_s) - \sum_{r \in R} d_r(f_r)
\text{ subject to } (\ref{cons:supply-demand-eq}). \label{opt:WM}
\end{equation}

\newcommand{\node}{\nu}

\begin{example}
Consider a network of coupled electricity markets for a single time period, described by a directed graph $G = (M,F)$ where $M$ is the set of nodes (electricity at each node corresponds to goods) and $F$ is the set of pairs of nodes (transmission parameters) that are connected by an edge. 
Electricity at each node $v \in M$ is encoded as a distinct good. As there are only capacity constraints on the flow, we do not require any additional flow parameters on top of the set of edges.
Each buyer $b \in B$ and seller $s \in S$ participates respectively in market $\node(b)$ and $\node(s)$. Meanwhile, for each directed edge $(v,w)$, there exists a transmission operator which may operate the line to transmit power between $v$ and $w$. Therefore, we parameterize the feasible bundles of transmission operators via the directional flow $f_{(v,w)}$. A feasible bundle provides a quantity $f_{(v,w)}$ of good $w$ and a quantity $-f_{(v,w)}$ of good $v$, and the cost for this flow is denoted $d_{(v,w)}(f_{(v,w)})$. We assume that line flows are only subject to capacity constraints,
$$ \underline{F}_{(v,w)} \leq f_{(v,w)} \leq \overline{F}_{(v,w)} \ \forall \ (v,w) \in F. $$
We therefore set $d_{(v,w)}(f_{(v,w)}) = 0$ if $f_{(v,w)} \in [\underline{F}_{(v,w)}, \overline{F}_{(v,w)}]$ and $d_{{(v,w)}}(f_{(v,w)}) = +\infty$ otherwise. 

In this market the set of supply-demand constraints is given by
$$ \sum_{s \in S | \node(s) = v} y_s - \sum_{b \in B | \node(b) = v} x_b + \sum_{(w,v) \in F} f_{(w,v)} - \sum_{(v,w) \in F} f_{(v,w)} = 0 \ \forall \ v \in M,$$
and the matrix $B_r$ in (\ref{cons:supply-demand-eq}) is simply the oriented node-incidence matrix of $G$. An optimal allocation then prescribes buyers' consumption, sellers' supply and the flows on power lines such that the gains from trade are maximized.
\end{example}

Then, in a coupled market, \emph{prices} $p \in \mathbb{R}^{M \cup F}$ correspond to the per-unit cost of purchase of each unit of a good or flow. Utilities are assumed to be quasilinear in payment -- thus each market participant has utility
\begin{align}\label{def:utility}
	u_b(x|p) & = v_b(x) - p^T x && \ \forall \ b \in B,  \\
u_s(y|p) & = p^T y - c_s(y) && \ \forall \ s \in S, \nonumber \\
u_r(f|p) & = p^T B_r f - d_r(f) && \ \forall \ r \in R. \nonumber
\end{align} 
By optimizing over $x,y,f$ in (\ref{def:utility}), each market participant has an \emph{indirect utility function}, denoting the utility they would have from consuming / providing the bundle that maximizes their utility given prices,
\begin{align}
	\hat{u}_b(p) & = \max_x  v_b(x) - p^T x && \ \forall \ b \in B, \label{def:indirect-utility} \\
	\hat{u}_s(p) & = \max_y  p^T y - c_s(y) && \ \forall \ s \in S,  \nonumber \\
	\hat{u}_r(p) & = \max_f p^T B_r f - d_r(f) && \ \forall \ r \in R. \nonumber
\end{align}
By comparison with Definition \ref{def:conjugate}, we see that the indirect utility function is simply the convex conjugate of the preferences of buyers and sellers, $\hat{u}_b(p) = (-v_b)^*(-p)$ and $\hat{u}_s(p) = c^*_s(p)$ for any price vector $p$ and any buyer $b$ or seller $s$. Similarly for transmission operators, $\hat{u}_r(p) = d^*_r(B^T_r p)$.

An optimal allocation $(z^*_\ell)_{\ell \in L}$ and prices $p$ together are then said to form a \emph{Walrasian equilibrium} if the allocation of each market participant is utility-maximizing at the given prices.

\begin{definition}[Walrasian equilibrium in coupled markets] \label{def:WE}
	A price vector $p$ and a feasible allocation $(z^*_\ell)_{\ell \in L}$ in $\mathbb{R}^{M \cup F}$ form a \textit{Walrasian equilibrium} if:
	\begin{enumerate}
		\item (Market clearing) The supply-demand equivalence constraints (\ref{cons:supply-demand-eq}) are satisfied.
		\item (Envy-freeness) The allocation of every agent maximizes their utility at the prices -- i.e. for any market participant $\ell$, $u_\ell(z^*_\ell|p) = \hat{u}_\ell(p)$.
		\item (Budget balance) The sum of payments equals zero,
		$p^T \left( \sum_{s \in S} y^*_s - \sum_{b \in B} x^*_\ell + \sum_{r \in R} B_r f_r \right) = 0$.
	\end{enumerate}	
\end{definition}

At a Walrasian equilibrium, envy-freeness implies that no agent's utility can be less than that for consuming / supplying zero goods. Therefore, it is also \emph{individually rational} for each buyer and seller to participate in the market, as no agent earns a negative payoff as a result of their market participation. 
A market is \emph{convex} if $-v_b, c_s, d_r$ are all closed convex functions. This allows us to derive a version of the First and Second Welfare Theorem for coupled and convex markets.  

\begin{theorem}[The Welfare Theorems for Coupled and Convex Markets] 
Let price vector $p^* \in \mathbb{R}^{M \cup F}$ and the allocation  $(z_\ell)^*_{\ell \in L}$ be a Walrasian equilibrium, then this allocation maximizes social welfare. Conversely, if $(z_\ell)^*_{\ell \in L}$ is a welfare-maximizing allocation, then it can be supported by a Walrasian price vector $p^*$ that forms a Walrasian equilibrium.
\end{theorem}

\textit{Proof:} The theorem follows by first considering the \emph{welfare function}, defined as the value of the welfare maximization problem as its linear constraints are perturbed (cf. \citet{rockafellar2015convex} for a detailed discussion). Assuming strong supply-demand equivalence is required,\footnote{The discussion generalizes easily to the case of weak supply-demand equivalence} the social welfare $\omega : \mathbb{R}^{M \cup F} \rightarrow \extR$ is defined as a function of excess supply such that
\begin{align*}
\omega(\sigma) = \max_{x,y,f} & \ \sum_{b \in B} v_b(x_b) - \sum_{s \in S} c_s(y_s) - \sum_{r \in R} d_r(f_r) \\
\text{ subject to} & \ \sum_{s \in S} y_s - \sum_{b \in B} x_b + \sum_{r \in R} B_r f_r = \sigma.
\end{align*}
Thus $\omega$ is a convolution, and by Proposition \ref{prop:calculus}.4 the convex conjugate of $-\omega$ is given by 
\begin{equation*}
(-\omega)^*(p) = \sum_{b \in B} (-v_b)^*(p) + \sum_{s \in S} c^*_s(p) + \sum_{r \in R} d^*_r(B_r^T p) = \sum_{\ell \in L} \hat{u}_\ell(p).
\end{equation*}

Given that valuations are concave and costs are convex, $\omega$ is concave in all arguments and thus $-\omega$ is convex. As all constraints are linear, constraint qualification is satisfied and $-\omega$ is closed. Therefore, $-\omega$ admits a subdifferential $\partial (-\omega)(0)$ at $\sigma = 0$. Any element of the subdifferential provides prices that correspond to the per-unit value of the provision of an additional constraint violation. As the constraint fixes excess supply to $0$, this is precisely the value of the provision of an additional unit of each good to the market.

By the Fenchel-Young inequality, in general we have $(-\omega)^*(p) - \omega(0) \geq 0$, with equality holding if and only if $p \in \partial(-\omega)(0)$. Therefore, for an optimal solution $p$ of the \emph{subgradient problem} 
\begin{equation}
\min_p \sum_{\ell \in L} \hat{u}_\ell(p) - \omega(0), \label{opt:LDWM}
\end{equation}
we note that $p^T \left( \sum_{s \in S} y^*_s - \sum_{b \in B} x^*_b + \sum_{r \in R} B_r f_r \right) = 0 $ and $\omega(0) = \sum_{b \in B} v_b(x_b^*) - \sum_{s \in S} c_s(y_s^*) - \sum_{r \in R} d_r(f_r^*)$. Then by the Definition (\ref{def:utility}) of utilities and by re-arranging terms, this subgradient problem may be rewritten
\begin{align}
	& \min_p \sum_{\ell \in L} \hat{u}_\ell(p) - \omega(0) \nonumber \\
	= & \min_p \sum_{b \in B} \hat{u}_b(p) + p^T x^*_b - v_b(x^*_b) + \sum_{s \in S} \hat{u}_s(p) + c_s(y^*_s) - p^T y^*_s + \sum_{r \in R} \hat{u}_r(p) + d_r(f^*_r) - p^T B_r^T f^*_r \nonumber \\
	= & \min_p \sum_{\ell \in L} (\hat{u}_\ell(p) - u_\ell(z^*_\ell|p)). \label{opt:min-GLOC}
\end{align}
For any market participant $\ell$ and for any price vector $p$, $\hat{u}_\ell(p) \geq u_\ell(z^*_\ell|p)$. Therefore as the value of the problem equals zero, at an optimal solution $p$ these must all hold with equality, which implies that an optimal solution to (\ref{opt:LDWM}) $p$ together with a welfare-maximizing allocation $(z^*_\ell)_{\ell \in L}$ form a Walrasian equilibrium. Likewise, if $p,(z^*_\ell)_{\ell \in L}$ form a Walrasian equilibrium then the value of (\ref{opt:min-GLOC}) is zero, which implies that $(z^*_\ell)_{\ell \in L}$ maximizes social welfare. \qed

We do not need differentiability, and convexity of $-v_b$, $c_s$, $d_r$ is sufficient for this proof. As convex optimization is also computationally efficient so long as each valuation and cost function is tractable to compute, the subgradient provides a natural way of computing prices. This motivates us to define \textbf{dual pricing functions} for an optimal outcome $\left( (x^*_b)_{b \in B}, (y^*_s)_{s \in S}, (f^*_r)_{r \in R} \right)$,
\begin{equation}
\lambda_b(p|x^*_b) = \hat{u}_b(p) - u_b(x^*_b|p), \ \lambda_s(p|y^*_s) = \hat{u}_s(p) - u_s(y^*_s|p), \ \lambda_r(p|f^*_r) = \hat{u}_r(p) - u_r(f^*_r|p). \label{def:dual-pricing}
\end{equation}
Furthermore, we call (\ref{opt:LDWM}) the \textbf{dual pricing problem} associated with the welfare maximization problem (\ref{opt:WM}).

Although the connections between convexity and the existence of a Walrasian equilibrium are well-known \citep{bikhchandani1997competitive, liberopoulos2016critical}, the version of the welfare theorems for coupled markets clearly delineates when we can expect Walrasian prices in coupled markets. This provides a foundation for our discussion of pricing rules in non-convex electricity markets. The Lagrangian dual and subgradients can also be used to derive prices in non-convex markets. Due to a positive duality gap, however, such prices no longer imply a Walrasian equilibrium. 

Note that in this paper, we do not consider how different pricing rules incentivize participants to bid truthfully. We assume large markets as complete-information games and argue that individual participants have little market power and act as price takers.
\section{Pricing Non-Convex and Coupled Markets} \label{sec:Theory}

In the absence of convexity, the negative welfare function $-\omega$ is non-convex in general and the subgradient problem (\ref{opt:LDWM}) has value $> 0$, pointing to a duality gap. The representation (\ref{opt:min-GLOC}) of the convex pricing problem then suggests that for any welfare-maximizing outcome $(z^*_\ell)_{\ell \in L}$ and any price vector $p^*$, Definition \ref{def:WE}.2 is not satisfied and market participants incur \emph{lost opportunity costs} (LOCs). This raises the question of how to price these markets.

In this section, we review some proposals for pricing optimal outcomes of such non-convex markets via the formalism through which we established the existence of Walrasian equilibria in coupled and convex markets. We will see that these pricing rules consist of convex models for the dual pricing problem (\ref{opt:LDWM}). Moreover, they assert a convexified model of the original welfare maximization problem corresponding to minimization of a \emph{class} of LOCs, given prices, as a central design goal: \emph{global} LOCs (GLOCs) which correspond to unrealized profit as participants deviate to any feasible outcome, \emph{local} LOCs (LLOCs) which correspond to unrealized profit as participants deviate to another outcome under fixed commitment, and \emph{make-whole payments} (MWPs) as the required amount of compensation to market participants to ensure they do not make a loss.  

We conclude the section by illustrating that the minimization of each class of LOCs results in prices causing market participants to incur large LOCs in other classes. This motivates framing pricing in non-convex markets as a multi-objective optimization problem. 

\subsection{Minimizing Global Lost Opportunity Costs}

\textit{Global lost opportunity costs} quantify the incentives of market participants to deviate from the optimal allocation. In particular, they are defined as the difference between the maximum profit a market participant can achieve (obtained from the indirect utility function) and the actual profit they yield under the optimal allocation (obtained from the direct utility function), given some prices $p$. 
\begin{definition}
	Let $\ell \in L$ be a market participant with allocation $z^*_\ell$, facing prices $p \in \mathbb{R}^M$. Then the \textbf{(global) lost opportunity cost} (GLOC) of $\ell$ is given 
	\begin{equation}
		\text{GLOC}_\ell(p|z^*_\ell) = \hat{u}_\ell(p) - u_\ell(z^*_\ell|p).
	\end{equation}
\end{definition}
Zero GLOCs for every market participant would imply no incentives to deviate from the optimal solution and therefore a Walrasian equilibrium. 
Minimizing GLOCs aims to minimize envy and thus to approximate a Walrasian equilibrium as much as possible. 
As mentioned earlier, in highly regulated electricity markets, GLOCs are typically not compensated by side-payments. Instead, market operators impose penalties of at least GLOC$_\ell$ on each participant $\ell$ if they deviate from the optimal allocation. Thus envy-freeness of prices might be less of a concern in such transparent and regulated markets.

The respective pricing rule that minimizes GLOCs is known as Convex Hull pricing because the associated welfare maximization problem is obtained by convexifying the preferences of market participants in the original welfare maximization problem \citep{hogan2003minimum,gribik2007market}.
\begin{equation}
	\max_{x,y,f} \sum_{b \in B} \conch[v_b](x_b) - \sum_{s \in S} \ch[c_s](y_s) - \sum_{r \in R} \ch[d_r](f_r) 
	\text{ subject to }  (\ref{cons:supply-demand-eq}). \label{opt:CH-pricing}\tag{Primal CH}  
\end{equation}

Given the optimal allocation $(z^*_\ell)_{\ell \in L}$, we thus relabel the contribution $\lambda_\ell(p|z^*_\ell)$ of market participant $\ell \in L$ to the dual pricing problem as $\lambda^{CH}_\ell(p|z_\ell^*)$ and call this contribution the \textit{Convex Hull pricing function} of $\ell$ at $z^*_\ell$.
From Definition (\ref{def:subgradient}), we infer that if $p$ is a solution to (\ref{opt:LDWM}) with objective value $0$, then 
\begin{equation*}
	- p  \in \partial (-v_b)^{**}(x^*_b) \ \forall \ b \in B, \quad
	p \in \partial c_s^{**}(y^*_s) \ \forall \ s \in S,  \quad 
	B_r^T p  \in \partial d^{**}_r(f^*_r) \ \forall \ r \in R.
\end{equation*}
In other words, when the objective of the CH pricing problem equals zero, prices form supragradients of the concave closures of buyers' valuation functions and subgradients of the convex closure of sellers' and transmission operators' cost functions. Furthermore, in this case the duality gap equals zero, the concave closures at the optimal allocation equal exactly the valuation functions and the convex closures equal the cost functions. This implies that prices reflect the marginal valuations and costs each market participant has for any extra unit of a good or flow. In general, CH pricing provides a supragradient of the \emph{concave closure of the welfare function} at $\sigma=0$, though these prices might not necessarily belong to the subdifferential of market participants at an optimal outcome when there exists a positive duality gap due to non-convexities. 

One issue with CH prices is that they are generally intractable to compute \citep{Schiro.2016}. There have been efforts to establish conditions under which CH prices can be computed by simple linear programs \citep{Hua.2017} or to design more efficient algorithms \citep{Andrianesis.2021,Knueven.2022}, yet as of today CH pricing cannot be applied for practical problems. Certain approximations of the dual pricing problem have been proposed as simple heuristics, for example taking the continuous relaxation when the welfare maximization problem (\ref{opt:WM}) is a mixed-integer welfare maximization problem. Such an approximation expands the feasible region of the dual pricing problem, providing tractability by accounting for exaggerated lost opportunity costs during price computation. This is referred to as the \textit{ELMP} pricing rule \citep{MISO.2019}. Note that these approaches are only used to derive prices, i.e. after the allocation problem has been solved to optimality. 

Besides, CH prices have come under scrutiny due to their signaling properties. CH prices do not signal network congestion properly and they allow offline units to set prices \citep{Schiro.2016}. We demonstrate in our experimental results that CH prices lead to high MWPs and LLOCs, and thus there might be price differences among adjacent network nodes, even though there is no congestion on the link connecting both. 

\subsection{Minimizing Local Lost Opportunity Costs}

\textit{Local lost opportunity costs} quantify the incentives of a market participants to deviate from the optimal allocation \textit{locally}. While we provide a more formal definition below, intuitively local lost opportunity costs consider only deviations which keep fixed the commitment status of the optimal allocation. For example, if a generator is committed in the optimal allocation, we do not consider lost opportunity costs that result from uncommitting at a certain time period, or leaving the market altogether. We will also show that minimizing LLOCs coincides with the widely used IP pricing rule \citep{oneill2005efficient}. 

To discuss local lost opportunity costs formally, we first present the notion of a piecewise convex function. The disjoint convex sets which form the domain of each market participant's cost or valuation functions may be interpreted as different commitment levels / operating ranges.

\begin{definition}
	We say that a function $f : \mathbb{R}^n \rightarrow \extR$ is \textit{piecewise convex} if there exist disjoint convex sets $X_1, X_2,.., X_{K} \subseteq \mathbb{R}^n$ and closed convex functions $f_1, f_2, .., f_k$ such that for any $1 \leq k \leq K, f_k : X_k \rightarrow \mathbb{R}$ and $f = \min_{1 \leq k \leq K} f_k$. For each $k \in K$, we let $\dom(f_k) = X_k$ denote the \emph{domain} of $f_k$. For any $x \in \mathbb{R}^n$, $f_{k'}$ is called \textit{active} at $x$ if $k' \in \arg \min_{1 \leq k \leq K} f_k(x)$.
\end{definition}

In particular, most U.S. and European electricity markets implement bidding languages which allow the expression of piecewise-concave valuations for buyers and piecewise-convex costs for sellers. This is because, in electricity markets, whether a market participant is committed or not is encoded via binary variables.

\begin{example}\label{ex:piecewise-convex}
	Consider, in a market with a single good, a generator $s$ with a cost function 
	\begin{align*}
		c_s(y) & = \min_u c y + h u \textnormal{ subject to } \underline{P} \cdot u \leq y \leq \overline{P} \cdot u, u \in \{0,1\} \\
		& = \begin{cases}
			0 & y = 0, \\
			cy + h & \underline{P} \leq y \leq \overline{P}, \\
			+\infty & \textnormal{else.}
		\end{cases}
	\end{align*}
	This is in fact the cost function of a generator with marginal cost $c$ and fixed cost $h$ for being committed. The cost function is in fact the minimum of two convex functions $c'_s, c''_s$ and is thus piecewise convex,
	$$c'_s(y) = \begin{cases}
		0 & y = 0 \\
		+\infty & y \neq 0
	\end{cases}, \quad\quad c''_s(y) = \begin{cases}
		cy + hu & \underline{P} \leq y \leq \overline{P} \\
		+\infty & \textnormal{else}
	\end{cases} .$$
\end{example}

The definition of piecewise convex functions thus motivates us to define local lost opportunity costs as the difference between the maximum profit a market participant may achieve amongst allocations in the domain of their active cost function, and the actual profit they obtain under the optimal allocation, given prices $p$.

\begin{definition}
	Let $\ell \in L$ be a market participant with a piecewise-convex cost function, who obtains allocation $z^*_\ell$ and faces prices $p \in \mathbb{R}^M$. Suppose that $c'$ is the active cost function of $\ell$, and write
	\begin{equation}
		\hat{u}'_\ell(p) = \max_z p^T z - c'_\ell(z)
	\end{equation}
	for the indirect utility function of $\ell$ with respect to their active cost function. Then the \textbf{local lost opportunity cost} (LLOC) of $\ell$ is given 
	\begin{equation}
		\text{LLOC}_\ell(p|z^*_\ell) = \hat{u}'_\ell(p) - u_\ell(z^*_\ell|p).
	\end{equation}
\end{definition}

Prices that imply zero LLOCs are often considered to generalize marginal pricing for non-convex markets as prices equal the marginal value of electricity. This is because when we restrict attention to only active cost / valuation functions, we obtain a convex market for which Walrasian prices are marginal costs. In fact, the widely used IP pricing rule \citep{oneill2005efficient} minimizes LLOCs. As in Example \ref{ex:piecewise-convex}, 
piecewise-convex preferences may be modeled via the addition of binary variables that represent the choice of active valuation / cost functions. When each $v_{b}, c_{s}, d_{r}$ is also piecewise linear, this allows casting the welfare maximization problem as a mixed-integer linear program. It is shown in \citet{oneill2005efficient} that marginal prices which eliminate LLOCs can be obtained by fixing binary variables to their optimal values and solving the resulting dual problem. This provides what is known as IP pricing in electricity markets.

Formally, IP pricing is given by the dual pricing problem associated with the welfare maximization problem
\begin{equation}
	\max_{x,y,f} \sum_{b \in B} \hat{v}_b(x_b) - \sum_{s \in S} \hat{c}_s(y_s) - \sum_{r \in R} \hat{d}_r(f_r) 
	\text{ subject to }  (\ref{cons:supply-demand-eq}), \label{opt:IP-pricing-primal} \tag{Primal IP}  
\end{equation}
where $\hat{v}_b, \hat{c}_s, \hat{d}_r$ are the corresponding active valuation and cost functions given the optimal allocation. The dual pricing problem can be equivalently stated as the minimization of LLOCs of all market participants,
\begin{align} \min_{p} & \ 
 \sum_{b \in B} \max_{x_b} \hat{v}_b(x_b) - p^T x_b - u_b(x^*_b|p) 
+ \sum_{s \in S} \max_{y_s} p^T y_s - \hat{c}_s(y_s) - u_s(y^*_s|p) \label{opt:IP-pricing} \tag{IP Pricing} \\
& \ \ 
+  \sum_{r \in R} \max_{f_r} p^T B_r f_r - \hat{d}_r(f_r) - u_r(f^*_r|p) \nonumber \\ 
\text{subject to} & \quad \ x_b \in \dom(\hat{v}_b) \ \forall \ b \in B, \quad y_s \in \dom(\hat{c}_s) \ \forall \ s \in S, \quad f_r \in \dom(\hat{d}_r) \ \forall \ r \in R,  \nonumber \end{align} 
minimizing the total LLOCs incurred by market participants for deviations under fixed commitment. We then denote the contribution of market participant $\ell$ to the dual pricing problem as $\lambda^{IP}_\ell(p|z_\ell^*)$ and call this contribution the \textit{IP dual pricing function} of $\ell$ at $z^*_\ell$.

It is important to note that LLOCs are an important indicator for inadequate congestion signals. With fixed commitment, prices also reflect the marginal value of additional transmission and price differences across nodes only occur when the network is congested, i.e. transmission network cannot allow for further profitable trades between the nodes. Therefore IP pricing provides adequate signaling of congestion in the network as prices reflect the marginal value of additional transmission capacity \citep{Yang.2019}, mitigating a shortfall of congestion income that was identified for CH pricing. On the other hand, in the presence of non-convexities CH pricing may result in large LLOCs (cf. Section \ref{sec:conflict}). In this case, market participants have high incentives for just small deviations from the welfare-maximizing allocation. Furthermore, the resulting incorrect congestion signals might also set flawed incentives for market participants, e.g., for demand response. 

\subsection{Minimizing Make-Whole Payments}
\textit{Make-whole payments} quantify the losses that a market participant incurs under the optimal allocation, given some prices $p$. This is equivalent to lost opportunity costs that only consider the deviation to non-participation, i.e. leaving the market entirely. 
\begin{definition}
	Let $\ell \in L$ be a market participant with allocation $z^*_\ell$, facing prices $p \in \mathbb{R}^M$. Then the \textbf{make-whole payment} required for $\ell$ is given,
\begin{equation}
	\text{MWP}_\ell(p|z^*_\ell) = \max\{-u_\ell(z^*_\ell|p),0\}.
\end{equation} 
\end{definition}
Market participants are compensated for their MWPs, resulting in a revenue imbalance for the market operator. Usually, MWPs are funded by additional charges imposed on consumers. As discussed in the introduction, increasing MWPs pose a concern in electricity markets, implying discriminatory pricing and diminishing the signaling value of prices. For IP prices, some market participants can incur substantial losses, as we find in our numerical experiments. 

To rectify this issue, \citet{Bichler.2021} introduced the following optimization problem, where MWPs are minimized directly.
\begin{align}
	\min_{\lambda \geq 0, p,\gamma} & & \sum_{b \in B} \lambda_b^{MWP}(p|x^*_b) + \sum_{s \in S} \lambda_s^{MWP}(p|y^*_s) + \sum_{r \in R} \lambda_r^{MWP}(p|f^*_r) &  \label{opt:min-MWP} \tag{Min-MWP}\\
	\text{subject to} & & - v_b(x^*_b) + p^T x^*_b - \lambda_b^{MWP}(p|x^*_b) & \leq 0 \ \forall \ b \in B, \nonumber \\
	& &  -p^T y^*_s + c_s(y^*_s) - \lambda_s^{MWP}(p|y^*_s) & \leq 0 \ \forall \ s \in S, \nonumber \\
	& &  -p^T B_r f^*_r  + d_r(f^*_r) - \lambda_r^{MWP}(p|f^*_r) & \leq 0 \ \forall \ r \in R . \nonumber
\end{align}
The associated dual pricing functions for this problem are $\lambda_\ell^{MWP}(p|z^*_\ell) = \max\{-u_\ell(z^*_\ell|p) , 0  \}$ for each market participant $\ell$, and we call this contribution the \textit{Min-MWP dual pricing function} of $\ell$. By definition the Min-MWP dual pricing function accounts only for the lost opportunity cost with respect to non-participation, thus the value of (\ref{opt:min-MWP}) is precisely the minimum MWPs required to compensate participants' losses. 

\subsection{Conflicting Design Goals}\label{sec:conflict}

We established in the preceding discussion that CH pricing, IP pricing, and Min-MWP pricing each optimize a corresponding class of LOCs. In practice, typically only MWPs are paid out to market participants, while all remaining LOCs are enforced by penalizing market participants for deviating from the optimal dispatch. However, regardless of the payment rule, the minimization of all LOC classes is evidently desirable. Yet as we demonstrate in the following, focusing on only one such design goal by using the pricing rules introduced above can lead to large LOCs in other classes. This finding follows from the different objective functions, but we argue that trade-offs can be substantial and that a pricing rule should thus consider all classes of LOCs, as the following example shows. 

\begin{example}\label{ex:CH-pricing}
	Consider a single-item market with two sellers in a single hour and two coupled locations. The first (second) seller is located at the first (second) node and has a minimum sales quantity of 2 (8) units, a maximum quantity of 15 (15) units, variable per-unit costs of \$10 (\$1) and fixed costs of \$1000 (\$10) when supplying a positive amount. At the first (second) node a fixed demand of 3 (1) units needs to be satisfied, and the line capacity is 2 in either direction. The optimal solution is for the first seller to supply the entire demand. In this case, $1$ unit is transmitted from the first node to the second node, and there is no congestion in the network. We record prices, GLOCs, MWPs, LLOCs for the pricing rules under consideration in Table \ref{tab:smallExI}. While CH pricing minimizes GLOCs, it may very well imply large MWPs and LLOCs and a false congestion signal (by the price difference across the uncongested line).
	\ifbool{tablesInText}{
	\begin{table}[H]
		\centering
		\begin{tabular}{c|cc|ccc}
			& Price 1 [\$/Unit] & Price 2 [\$/Unit] & GLOCs [\$] & MWPs [\$] & LLOCs [\$] \\
			\hline 
			CH & 76.67 & 1.67 & 733.33 & 733.33 & 733.33  \\
			IP & 10.00 & 10.00 & 1125.00 & 1000.00 & 0 \\
			Min-MWP & 260.00 & 260.00 & 6625.00 & 0 & 2750.00
		\end{tabular}
		\caption{Example for CH pricing with large MWPs and LLOCs}
		\label{tab:smallExI}
	\end{table}
	} {}
	To illustrate the reason for this discrepancy, we compute participants' preferences and the optimal allocation for (\ref{opt:CH-pricing}). By convexification the seller's minimum loads are relaxed and no longer binding. As a result, the first (second) seller can supply up to 15 units with variable per-unit costs of \$$10 + \frac{1000}{15}$ (\$$1 + \frac{10}{15}$), while the fixed demand and line capacity are unchanged. Therefore, the (hypothetical) optimal allocation is for the second seller to provide 3 units while the first seller provides 1 unit, and 2 units are transmitted from the second node to the first. The first seller's marginal costs sets the price at the first node, and the second seller's marginal costs sets the price at the second node. However, in the true optimal allocation the second seller could not be committed (due to its minimum quantity of 8), and 1 unit was transmitted in the opposite direction.
\end{example}

In contrast to CH pricing, IP prices fix LLOCs to zero via local convexity, but discard any information on changing commitment. As a result, IP prices may fail to adequately account for GLOCs or MWPs. 

\begin{example}
	Consider a single-item market with two sellers in a single hour and two coupled locations. The first (second) seller is located at the first (second) node and has a minimum sales quantity of 2 (8) units, a maximum quantity of 8 (15) units, variable per-unit costs of \$1 (\$10) and fixed costs of \$100 (\$100) when supplying a positive amount. At the first (second) node a fixed demand of 6 (1) units needs to be satisfied, and the line capacity is 4 in either direction. The optimal solution is for the first seller to supply the entire demand, and there is no congestion in the network. We record prices, GLOCs, MWPs, LLOCs for different pricing rules in Table \ref{tab:smallExIII}. While IP pricing minimizes LLOCs, it may very well imply large GLOCs and MWPs.
	\ifbool{tablesInText}{
	\begin{table}[H]
		\centering
		\begin{tabular}{c|cc|ccc}
			& Price 1 [\$/Unit] & Price 2 [\$/Unit] & GLOCs [\$] & MWPs [\$] & LLOCs [\$] \\
			\hline 
			CH & 13.5 & 13.5 & 12.5 & 12.5 & 12.5 \\
			IP & 1 & 1 & 100 & 100 & 0 \\
			Min-MWP & 15.29 & 15.29 & 14.29 & 0 & 14.29
		\end{tabular}
		\caption{Example for IP pricing with large GLOCs and MWPs}
		\label{tab:smallExIII}
	\end{table}
	} {}
\end{example}

Finally, Min-MWP directly minimizes the make-whole payments but loses all information on subgradients in the process. As a result, while MWPs may be reduced to insignificant amounts, the solution set will be large and the computed prices will in general incur both significant LLOCs and GLOCs.

\begin{example}
	Consider a single-item market with two sellers in a single hour and two coupled locations. The first (second) seller is located at the first (second) node and has a minimum sales quantity of 2 (8) units, a maximum quantity of 50 (15) units, variable per-unit costs of \$10 (\$10) and fixed costs of \$1000 (\$10) when supplying a positive amount. At the first node there is a fixed demand of 4 units, at the second node a buyer is willing to pay \$50 per unit to consume up to 3 units, and the line capacity is 2 in either direction. The optimal solution is for the first seller to supply the entire demand at the first node and 2 units to the second node, and as a consequence there is congestion in the network. We record prices, GLOCs, MWPs, LLOCs for different pricing rules in Table \ref{tab:smallExII}. While Min-MWP pricing minimizes MWPs, it may very well imply large GLOCs and LLOCs and the congestion is not reflected in the prices as the price is identical at both nodes. In contrast, with IP pricing the price on the second node is higher than on the first node, indicating correctly directed congestion.
	\ifbool{tablesInText}{
	\begin{table}[H]
		\centering
		\begin{tabular}{c|cc|ccc}
			& Price 1 [\$/Unit] & Price 2 [\$/Unit] & GLOCs [\$] & MWPs [\$] & LLOCs [\$] \\
			\hline 
			CH & 30.00 & 10.67 & 996.67 & 918.67 & 996.67 \\
			IP & 10.00 & 50.00 & 1590.00 & 1000.00 & 0 \\
			Min-MWP & 176.67 & 176.67 & 10076.67 & 253.33 & 7586.67
		\end{tabular}
		\caption{Example for Min-MWP pricing with large GLOCs and LLOCs}
		\label{tab:smallExII}
	\end{table}
	}{}
\end{example}

The examples in this section might appear artificial to the reader. However, we remark that our numerical results verify that this contradiction between design objectives persists when we consider larger and more realistic market data in Section \ref{sec:Electricity}. 

\section{Pricing as Multi-Objective Optimization} \label{sec:combinations}

In the previous section, we discussed dual pricing problems that minimize either GLOCs (CH pricing), LLOCs (IP pricing), or MWPs (Min-MWP pricing). We provided examples illustrating that a focus on one of these objectives can lead to high costs in the others. 
Ultimately, we are faced with trade-offs between the three conflicting objectives. We note that the market designer can also have design goals that go beyond minimization of deviation incentives, e.g. (approximate) budget-balance. However, we argue that a pricing rule that finds a good trade-off between GLOCs, LLOCs, and MWPs already satisfies crucial requirements for electricity markets: low incentives to deviate from the optimal allocation, adequate congestion signals, as well as low side-payments that distort the price signal. We therefore infer that the pricing problem in non-convex markets is one of multi-objective optimization rather than single-objective optimization, which deviates from prior literature on the subject.

While many techniques for multi-objective optimization have been proposed (see, e.g., \citet{Deb.2001}, \citet{miettinen2012nonlinear} or \citet{Emmerich.2018}), not all of them are suitable to be used as pricing rules in electricity markets. 
A poor choice of a multi-objective (dual pricing) formulation can impose a highly distorted (primal) convex model for the welfare maximization problem, resulting in a divergence between its implied optimal allocation and the true optimal allocation. For example, the primal convex model implied by the multi-objective formulation might possess fictitious goods and agents, unrealistically expand the feasible sets of market participants, or consolidate market participants to a single entity. This is a subtle issue that we refer to as \emph{economic interpretability problem}, and discuss in detail in Appendix \ref{app:economic_interpretability}. 
Observe that the introduced dual pricing problems possess an additively separable objective function with linear prices for each good, i.e. they are of the form $\min_{p \in \mathbb{R}^M} \sum_{\ell \in L} \lambda_\ell(p|z^*_\ell)$.
We require the same from alternative pricing rules and respective multi-objective optimization problems, in order to prevent distortions in the number of market participants and goods, and thus to allow for economic interpretability.
Each dual pricing function is a representation of some lost opportunity cost terms. Thus, we consider multi-objective problems which account for trade-offs between GLOCs, LLOCs, and MWPs. Linear scalarization is the most well-known multi-objective optimization technique and it satisfies the above requirement. Therefore, we first discuss linear scalarization before we introduce a new multi-objective optimization technique that also satisfies separability but is parameter-free: the join.

\subsection{Linear Scalarization}

Linear scalarization optimizes a weighted sum of individual objective functions. In particular, given dual pricing functions for market participants $\left(\sum_{\ell \in L} \lambda^i_\ell(\cdot | z^*_\ell)\right)_{i \in \{1,..,m\}}$ associated with $m$ dual pricing problems and a non-negative weight vector $w \in \mathbb{R}^m_{\geq 0}$, the linear scalarization problem is given by

\begin{equation}
	\min_{p} \sum_{\ell \in L} \sum_{i=1}^m w_i \cdot \lambda^i_\ell(p| z^*_\ell) \label{opt:LS}
\end{equation}

As each dual pricing objective $\lambda^i$ is separable over the market participants, linear scalarization is also equivalent to picking certain convex approximations to each participant's valuation or cost function, as in the case for CH, IP, and Min-MWP pricing. In fact, the resulting pricing problem is that of a convex market where each participant is replaced by an \emph{admixture} of participants, where preferences from each dual pricing problem $i$ are present in proportion to the weight $w_i$. Specifically, suppose the weights are non-negative and normalized such that $\sum_{i = 1}^m w_i = 1$. Then, if buyer $b$ has allocation $x^*_b$ and valuation function $v^i_b$ in the welfare maximization problem corresponding to dual pricing objective $i$, by Proposition \ref{prop:calculus} the weighted dual pricing function $\sum_{i = 1}^m w_i \lambda^i(p| x^*_b)$ corresponds to the valuation function
\begin{equation}
\bar{v}_b(x) = \max \sum_{i = 1}^m w_i v^i_b ( \chi^i_b ) \text{ subject to } \sum_{i = 1}^m w_i \chi^i_b = x. \label{conv:buyer}
\end{equation}
The prices computed via linear scalarization then equal the marginal value of electricity for this mixed market. 

In a multi-objective optimization problem, it is generally not possible for a solution to be optimal for each objective. Therefore, the optimality criterion is, in general, \emph{Pareto optimality}. For $m$ dual pricing problems, a price vector $p$ is said to be \textbf{Pareto optimal with respect to objectives} if there does not exist another price vector $q$ such that for any pricing problem $i$, $\sum_{\ell \in L} \lambda_\ell^i(q|z^*_\ell) \leq \sum_{\ell \in L} \lambda_\ell^i(p|z^*_\ell)$, with one inequality holding strictly.

It is known that solutions to the linear scalarization problem yield the Pareto frontier of the multi-objective \citep{Emmerich.2018}. We thus infer that if the dual pricing functions under consideration correspond to CH, IP, and Min-MWP pricing, linear scalarization allows to obtain other Pareto-efficient solutions with respect to total GLOCs, MWPs, and LLOCs. 

However, linear scalarization comes with certain disadvantages for practical applications despite providing a Pareto optimal solution with respect to GLOCs, MWPs, and LLOCs. In particular, it requires preference information in order to set weights that produce a desirable outcome. In practice, however, preferences might be difficult to define in an environment with various stakeholders and without being able to study the impact of weights on the outcomes. Market operators might want to re-calibrate and fine-tune the weights after the Pareto frontier has been computed. For this reason, linear scalarization cannot be considered a full a priori multi-objective technique as weights and therefore the pricing rule might be set only after bids have been elicited. Therefore, linear scalarization would be difficult to implement in practice.

\subsection{Join}

In what follows, we focus on a pricing rule that treats low LLOCs and low MWPs as first-order objectives. Together, they imply that the market allocation is as locally stable as possible against considerations of non-participation or small deviations by the market participants. While GLOCs are not accounted for, we will see as a result in our experiments that GLOCs remain comparable to IP pricing. 

Given prices $p$ and an optimal allocation $(z^*_\ell)_{\ell \in L}$, market participant $\ell$'s opportunity cost under consideration of local deviations in an allocation is given by $\lambda_\ell^{IP}(p|z^*_\ell)$ while their opportunity cost against exiting the market is given by $\lambda^{MWP}_\ell(p|z^*_\ell)$. Therefore, the amount of compensation required to disincentivize such deviations from $\ell$ is given by $\max \{ \lambda^{IP}_\ell(p|z^*_\ell), \lambda^{MWP}_\ell(p|z^*_\ell)\}$. This motivates us to consider the following dual pricing problem.

\begin{definition}
	For each market participant $\ell$, let $\lambda_\ell^{IP}(p|z^*_\ell)$ and $\lambda_\ell^{MWP}(p|z^*_\ell)$ denote the LLOCs and MWPs of $\ell$ at prices $p$ and allocation $z^*$, respectively. Then the \textbf{join} IP $\lor$ MWP of IP and Min-MWP pricing is the dual pricing problem
	\begin{equation}
		\min_{p} \sum_{\ell \in L} \max\{\lambda_\ell^{IP}, \lambda_\ell^{MWP} \}(p|z^*_\ell).  \label{opt:join} \tag{IP $\lor$ MWP}
	\end{equation}
\end{definition}

By Proposition \ref{prop:calculus}.5, if $f,g$ are convex functions, then the convex conjugate of $\max\{f,g\}$ is the function $\ch \min \{f^*, g^*\}$. Thus (\ref{opt:join}) utilizes minimal concave closures of valuation functions and convex closures of cost functions that account for both LLOCs and MWPs. Note that if we considered $\lambda^{CH}_\ell(p|z^*_\ell)$ as well, the resulting pricing rule would correspond to the minimization of GLOCs and thus to CH pricing. This is because GLOCs capture all incentives to deviate, including local deviations and non-participation, and therefore are always greater or equal to both MWPs and LLOCs.
We also remark that when the welfare maximization problem is a MILP, the join prices may be found via a linear program. For instance, for the formulation of IP pricing in Appendix \ref{app:IP-Dual} and Min-MWP pricing in Appendix \ref{app:MWP-Dual}, the linear programming formulation of the join pricing problem is as shown in Appendix \ref{app:Join}. 

The main motivation to use join prices is based on the joint minimization of LLOCs and MWPs. In fact, the join IP $\lor$ MWP is guaranteed to achieve lower MWPs than IP pricing.

\begin{proposition}\label{prop:join-less-LLOC}
Suppose that $p^{\lor}$ is an optimal solution of (\ref{opt:join}) and $p^{IP}$ is an optimal solution of (\ref{opt:IP-pricing}). Then $\sum_{\ell \in L} \lambda^{MWP}_\ell(p^{\lor}|z^*_\ell) \leq \sum_{\ell \in L} \lambda^{MWP}_\ell(p^{IP}|z^*_\ell)$.
\end{proposition}

\textit{Proof:} The result follows since
\begin{align*}
\sum_{\ell \in L} \lambda^{MWP}_\ell(p^{\lor}|z^*_\ell) 
&\leq \sum_{\ell \in L} \max \{ \lambda^{MWP}_\ell , \lambda^{IP}_\ell \}(p^{\lor}|z^*_\ell) \\
& \leq \sum_{\ell \in L} \max \{ \lambda^{MWP}_\ell , \lambda^{IP}_\ell \}(p^{IP}|z^*_\ell) 
= \sum_{\ell \in L} \lambda^{MWP}_\ell (p^{IP}|z^*_\ell).
\end{align*}
Here, the first inequality holds since the terms of the second sum are element-wise no less than the terms of the first sum, the second inequality holds by the minimum property of $p^{\lor}$. The final equality then holds since for any market participant $\ell$, prices $p^{IP}$ achieve zero LLOCs due to convexity.
\qed

While Proposition \ref{prop:join-less-LLOC} might appear to immediately follow from its definition, this is only guaranteed since IP pricing achieves zero LLOCs for each market participant. Trying to modify the proof to show that a solution of (\ref{opt:join}) achieves lower LLOCs than any solution of (\ref{opt:min-MWP}), we obtain the following guarantee.

\begin{corollary}\label{cor:join-less-MWP}
	Suppose that $p^{\lor}$ is an optimal solution of (\ref{opt:join}) and $p^{MWP}$ is an optimal solution of (\ref{opt:min-MWP}) such that $\sum_{\ell \in L} \lambda_\ell^{MWP}(p^{MWP}|z^*_\ell) = 0$. Then $\sum_{\ell \in L} \lambda^{IP}_\ell(p^{\lor}|z^*_\ell) \leq \sum_{\ell \in L} \lambda^{IP}_\ell(p^{MWP}|z^*_\ell)$.
\end{corollary}

Therefore, a solution of (\ref{opt:join}) achieves lower LLOCs than any solution of (\ref{opt:min-MWP}) in settings where zero MWPs may be achieved with linear and anonymous prices. In general, the presence of non-convexities might render this impossible. Indeed, the following example demonstrates that the zero MWP condition in Corollary \ref{cor:join-less-MWP} is necessary.
\begin{example}
	Consider a market with one good, one seller $s$ and two buyers $b_1$ and $b_2$. Buyer $b_1$ has a block bid of $\$20$ for $1$ MWh and buyer $b_2$ has a block bid of $\$10$ for $1$ MWh, both of which must be fulfilled completely or not at all. Seller $s$ has a cost function for generation 
	$$ c_s(y) = \begin{cases}
		0 & y = 0, \\
		\frac{30}{\epsilon} \cdot (y - 2 + \epsilon) & y \in [2-\epsilon,2], \\
		+\infty & \textnormal{otherwise},
	\end{cases}
	$$
	with some small $\epsilon > 0$. In this case, note that only the seller $s$ can have strictly positive LLOCs, thus she determines IP prices and LLOCs are minimized for  $p^{IP} \geq 30/\epsilon$. Meanwhile, setting the price $p^{MWP}$ equal to $15$ minimizes MWPs. However, since $p^{\lor} \in [0,10]$, seller $s$ incurs a greater LLOC than for $p^{MWP}$.
\end{example}

A sufficient condition to satisfy the zero MWP condition with linear and anonymous prices is the presence of purely inelastic demand \citep{Bichler.2021}. 
In our numerical analysis in Section \ref{sec:Electricity} we also find that the set of optimal solutions to (\ref{opt:min-MWP}) is large and that a solver typically picks solutions with very high GLOCs and LLOCs. Then, the solution of (\ref{opt:join}) achieves lower LLOCs than the solution of (\ref{opt:min-MWP}) picked by the solver.
The join does not necessarily exhibit Pareto optimality with respect to the objectives of total LLOCs and total MWPs. However, the join satisfies a \emph{participant-wise} Pareto optimality criterion.

\begin{proposition}\label{prop:join-pareto}
	For an optimal outcome $(z^*_\ell)$ of a market, there exists some optimal solution $p^\lor$ of (\ref{opt:join}) such that deviations cannot jointly reduce MWPs and LLOCs of all market participants. Formally, there is no other price vector $q$, such that for any market participant $\ell$ and any $i \in \{IP, MWP\}$, $\lambda_\ell^{i}(q|z^*_\ell) \leq \lambda_\ell^i(p^\lor|z^*_\ell)$, with the inequality holding strictly for some $\ell, i$.
\end{proposition}

CH pricing dominates the literature on electricity market pricing. By minimization of GLOCs, CH prices naturally satisfy Pareto optimality. 
However, \citet{Schiro.2016} show that CH pricing allows offline sellers to distort both locational prices and congestion signals, as we have also demonstrated. 
The join avoids such distortions to the welfare maximization problem exhibited by CH pricing. 
It only modifies the dual pricing function of online units and thus disallows offline units from having any effect on the price. 
Unlike CH pricing, the join can also be implemented efficiently (in poly-time) in practical electricity markets. 
Overall, the join provides a straightforward, \textit{parameter-free} method of jointly minimizing LLOCs and make-whole payments, eliminating the need for fine-tuning or preference elicitation.

\section{Numerical Results} \label{sec:Electricity}

To analyze the proposed pricing rules in an exemplary electricity market, we consider a simplified multi-period \textit{direct current optimal power flow} (DCOPF) model \citep{Frank.2012, Molzahn.2019}. This model describes a nodal electricity market with linearized transmission flows.\footnote{Tighter power flow relaxations exist, but are currently not applied in practical electricity markets for computational reasons. We refer to \citet{Molzahn.2019} for a comprehensive overview of power flow problems.}
We refer to Appendix \ref{app:DCOPF_Notation} for the notation and problem formulation. The electricity market consists of a set of nodes $V$, with a specified reference node $R^* \in V$. Each buyer $b$ (seller $s$) has an associated node $\nu(b)$($\nu(s)$), and we denote by $N(v)$ the set of neighboring nodes to node $v$. With $T$ as the set of considered time periods and $B_{vw}$ as the line susceptances, we formulate the DCOPF as follows.

\begin{align}
	\max_{x,y,u,\phi,f,\alpha} & &\sum_{b \in B} v_b(x_b) - \sum_{s \in S} c_s(y_s, u_s, \phi_s) - \sum_{v \in V, w \in N(v)} d_{vw}(f_{vw}) - \sum_{v \in V} d_{v}(\alpha_{v})  & \label{opt:DCOPFprimal} \tag{DCOPF} \\
	\text{subject to} & &  \sum_{s \in S | \nu(s) = v} y_{st} - \sum_{b \in B | \nu(b) = v} x_{bt} + \sum_{w \in N(v)} B_{vw} (\alpha_{wt} - \alpha_{vt}) = 0 \quad \forall \ v \in V, t \in T, \nonumber \\ 
	& & f_{vwt} + B_{vw} \alpha_{wt} - B_{vw} \alpha_{vt} = 0 \quad \forall \ v \in V, w \in N(v), t \in T, \nonumber
\end{align}
where $x$ are variables corresponding to buyers' consumption at each time frame, $y$ are variables corresponding to the generation of sellers at each time frame, $u$ are binary variables that model non-convexities in sellers' cost functions, $\phi$ are binary variables that indicate generation start for sellers, $f$ are variables corresponding to flows on the grid, and $\alpha$ are variables that correspond to voltage angles.

To proceed with our analysis, we will impose an explicit form on our valuation and cost functions. We assume that the valuation and cost functions are given as follows:

\begin{enumerate}
\item For each buyer $b$, we have concave, piecewise linear valuation functions, represented by hourly bids $\beta_b^t$ for each time period $t$. Moreover, buyer $b$ has inelastic demand $\underline{P}_{bt}$, and has maximum power consumption $\overline{P}_{bt}$, 
\begin{align*}
v_b(x_b) = \max_{x_b} \sum_{t \in T, \ell \in \beta_b^t} x_{bt\ell} v_{bt\ell} & & \text{ subject to } & & x_{bt\ell} & \in [0,q_{bt\ell}]  &&\ \forall t \in T, \ell \in \beta_b^t, \\
&&&&x_{bt} - \sum_{\ell \in \beta_b^t} x_{bt\ell} & = \underline{P}_{bt} &&\ \forall t \in T, \\
&&&&x_{bt} & \leq \overline{P}_{bt} && \ \forall t \in T.
\end{align*}
\item For each seller $s$, when $u_{st} = 1$, seller $s$ has some convex cost function for producing a positive amount at time $t$ on some closed interval -- modeled again by hourly bids $\beta_s^t$. Moreover, $u_{st} \in \{0,1\}$ is a binary variable that denotes whether the unit of seller $s$ is active at time $t$ (associated with fixed costs $h_{s}$), and the variable $\phi_{st} = u_{st} - u_{s(t-1)}$ represents whether a unit is turned on or off at time $1 < t \leq T$. There potentially exists a set of minimum uptime constraints that are linear for $u_s, \phi_s$, 
\begin{align*}
c_s(y_s,u_s,\phi_s) = \min_{y_s,u_s,\phi_s} & & \sum_{t \in T, \ell \in \beta_s^t} y_{st\ell} c_{st\ell} & + \sum_{t \in T} h_{s} u_{st} & &  \\
\text{subject to} & & y_{st\ell} & \in [0,q_{st\ell}] \ \forall \ t \in T, \ell \in \beta_s^t,  & y_{st} - \underline{P}_{st} u_{st} & \geq 0 \ \forall \ t \in T, \\
& & u_{st} & \in \{0,1\} \ \forall \ t \in T,  & y_{st} - \overline{P}_{st} u_{st} & \leq 0 \ \forall \ t \in T, \\
& & \phi_{st} & \geq 0 \ \forall \ t \in T,  & \phi_{st} - u_{st} + u_{s(t-1)} & \geq 0 \ \forall \ 1 < t \leq T, \\
& & y_{st} - \sum_{\ell \in \beta_s^t} y_{st\ell} & = 0 \ \forall \ t \in T,  & \sum_{i = t - R_s + 1}^t \phi_{st} - u_{st} & \leq 0 \ \forall \ 1 < t \leq T.
\end{align*}
\item For the \ref{opt:DCOPFprimal}, transmission operators decide about line flows and phase angles. First, each flow between two nodes $v$ and $w$, $f_{vwt}$ lies on a closed interval $[\underline{F}_{vw}, \overline{F}_{vw}]$,
$$ d_{vw} (f_{vw}) = \begin{cases}
\infty & \exists 1 \leq t \leq T \text{ such that } f_{vwt} \notin [\underline{F}_{vw}, \overline{F}_{vw}], \\
0 & \text{else.}
\end{cases} $$
Second, phase angles $\alpha_{vt}$ at node $v$ can be set at zero cost. The phase angle at the reference node, however, is fixed to $0$ at each time period, 
$$ d_v ( \alpha_v ) = \begin{cases}
\infty & v = R^* \text{ and } \exists 1 \leq t \leq T, \alpha_{vt} \neq 0, \\
0 & \text{else.}
\end{cases}$$
\end{enumerate}

With this explicit welfare maximization problem at hand, we implement the pricing rules discussed in Section \ref{sec:Theory}. While CH pricing is intractable in general, a common heuristic in practical electricity markets is to employ ELMP \citep{MISO.2019}. ELMP relaxes the binary constraints in \ref{opt:DCOPFprimal} for $u_{st}$, $\{0,1\} \rightarrow [0,1]$, and prices are obtained from the resulting dual problem. It has been shown by \citet{Hua.2017} that with the valuation / cost functions in our setting, the dual of this relaxation of \ref{opt:DCOPFprimal} is in fact equivalent to CH pricing. Therefore, we have an explicit and tractable dual formulation of CH prices, provided in Appendix \ref{app:ELMP-Dual}.

In order to compute IP prices, we follow the approach suggested by \citet{oneill2005efficient}. In particular, we restrict each integer variable of the welfare maximization problem to the value it takes in the optimal allocation and solve the dual of the resulting problem. More explicitly, we set $u_{st}$ as $\{0,1\} \rightarrow [0,1]$ and $u_{st}=u_{st}^*$ in \ref{opt:DCOPFprimal} and solve the dual problem provided in Appendix \ref{app:IP-Dual}.

For our implementation of Min-MWP, we note that given some prices the transmission operators have either infinite or zero LLOCs (and hence GLOCs) associated with phase angles. To be able to obtain a solution to Min-MWP with finite LLOCs / GLOCs, we thus opt to replace the Min-MWP dual pricing functions for phase angles with their actual CH dual pricing functions in our implementation, thus accounting for their GLOCS. This implementation, provided in Appendix \ref{app:MWP-Dual}, minimizes MWPs of buyers, sellers and transmission operators under the constraint that no GLOCs occur with respect to phase angles.

For linear scalarization, we combine these pricing problems and optimize a weighted average of the individual objective functions. The formulation for the join (\ref{opt:join}) is provided in Appendix \ref{app:Join} and considers each participant's maximum of the IP and Min-MWP dual pricing function. 

We parameterize the \ref{opt:DCOPFprimal} for the IEEE RTS system (Section \ref{sec:rts}) as well as for the German bidding zone (Section \ref{subsec:bzr}) based on data that was recently used for the bid zone review in the European Union. In Appendix \ref{app:arpa}, we further report results for the ARPA-E Grid Optimization Competition.

\subsection{IEEE RTS System} \label{sec:rts}
First, we report results based on the IEEE RTS System, originally introduced by \citet{IEEERTSTaskForceofAPMSubcommittee.1979} and used in a variety of studies on electricity markets \citep{GarcaBertrand.2006, Morales.2009, Zoltowska.2016, Hytowitz.2020, Zocca.2021}. \citet{Grigg.1999} provide the stylized system topology, transmission network parameters, hourly (nodal) demand data as well as characteristics of generating units. In accordance with \citet{Zoltowska.2016}, we select the single-area, 24-node topology by \citet{Grigg.1999} for a representative 24-hour winter day with 32 generators (total capacity: 6.81 GW) and 17 consumers (average hourly demand: 2.60 GWh). Generators exhibit several non-convexities, such as no-load costs, minimum loads, or minimum runtimes. For data on generation costs or demand valuations we rely on the bid and offer curves provided by the case studies of \citet{GarcaBertrand.2006} and \citet{Zoltowska.2016} on this system. Generators and consumers are embedded in a DC power flow model with 24 nodes. The optimal dispatch is computed by solving the mixed-integer \ref{opt:DCOPFprimal} problem. All applied pricing rules constitute linear programs, with negligible computational effort compared to the initial DCOPF MILP. Table \ref{tab:ieee_stats} provides a first high-level overview of the results.\footnote{Note that Min-MWP has multiple optimal solutions, i.e. there exist multiple sets of prices (forming a polyhedron) that lead to zero MWPs. We report the solution that our solver provides.} 

\ifbool{tablesInText}{
\begin{table}[H]
	\centering
	\begin{tabular}{l|lll}
		& GLOCs [\$] & MWPs [\$] & LLOCs [\$]\\ \hline
		CH & 1436.21 & 202.57 & 1272.12 \\
		IP & 10364.24 & 3387.00 & 0 \\ 
		Min-MWP & 2.1 $\times 10^{11}$ & 0 & 1.1 $\times 10^{11}$ \\
		\hline
		0.9 IP + 0.1 Min-MWP & 11025.03 & 693.00 & 111.79 \\
		0.9 CH + 0.1 Min-MWP & 1432.93 & 199.27 & 1119.85 \\
		0.5 IP + 0.5 CH & 1570.13 & 606.21 & 729.29 \\ 
		\hline
		IP $\lor$ Min-MWP & 12022.38 & 0 & 191.14 
	\end{tabular}
	\caption{RTS System - GLOCs, MWPs, LLOCs}
	\label{tab:ieee_stats}
\end{table}
} {}

To begin with, the results underpin the finding that the optimization of individual objectives via CH, IP, and Min-MWP pricing is undesirable in at least one of the other objectives. 

\begin{result}
The LOC classes under consideration (GLOCs, LLOCs, MWPs) are conflicting. Pricing rules that optimize one class of LOCs lead to undesirable outcome in other classes of LOCs. 
\end{result}

Applying linear scalarization to IP and Min-MWP pricing leads to alternative results on the Pareto frontier (e.g., Table \ref{tab:ieee_stats} displays a linear scalarization with 90\% weight on IP and 10\% weight on Min-MWP\footnote{A simple 50:50 weight assignment would lead to a high emphasis on MWPs and outcomes that are much worse than that of the join.}), yet setting the weights is arbitrary. It should be noted that the set of outcomes implied by linear scalarization is not continuous. That is, there is only a limited number of outcomes and different weight vectors might produce the same result. The join IP $\lor$ Min-MWP requires no MWPs and LLOCs are reduced by 85\% compared to CH pricing (and close to 100\% compared to Min-MWP pricing). Moreover, GLOCs are significantly reduced compared to Min-MWP pricing and are only 16\% higher than the GLOCs of IP pricing.

Figure \ref{fig:ieee_3d} illustrates the GLOCs, MWPs, and LLOCs for the tested pricing rules and the Pareto frontier. Note that Min-MWP possesses very high GLOCs and LLOCs, and can therefore not be meaningfully depicted. The Pareto frontier is obtained by applying linear scalarizations to the three pricing rules and is presented in \ref{subfig:gloc_mwp_lloc}. The lines in Figures \ref{subfig:gloc_mwp} - \ref{subfig:mwp_lloc} correspond to pairwise linear scalarizations, i.e. projections of the Pareto frontier on the respective two-dimensional space. Those lines connect the discrete number of Pareto-efficient outcomes implied by linear scalarization.  

\begin{result}
	The Pareto frontier of GLOCs, LLOCs, and MWPs possesses a high curvature. 
	This suggests that there are significant merits to balancing the trade-offs between different classes of LOCs by applying multi-objective optimization. 
\end{result}

\ifbool{figuresInText}{
\begin{figure}[H]	
	\centering
	\begin{subfigure}{.475\linewidth}
		\includegraphics[width=\linewidth]{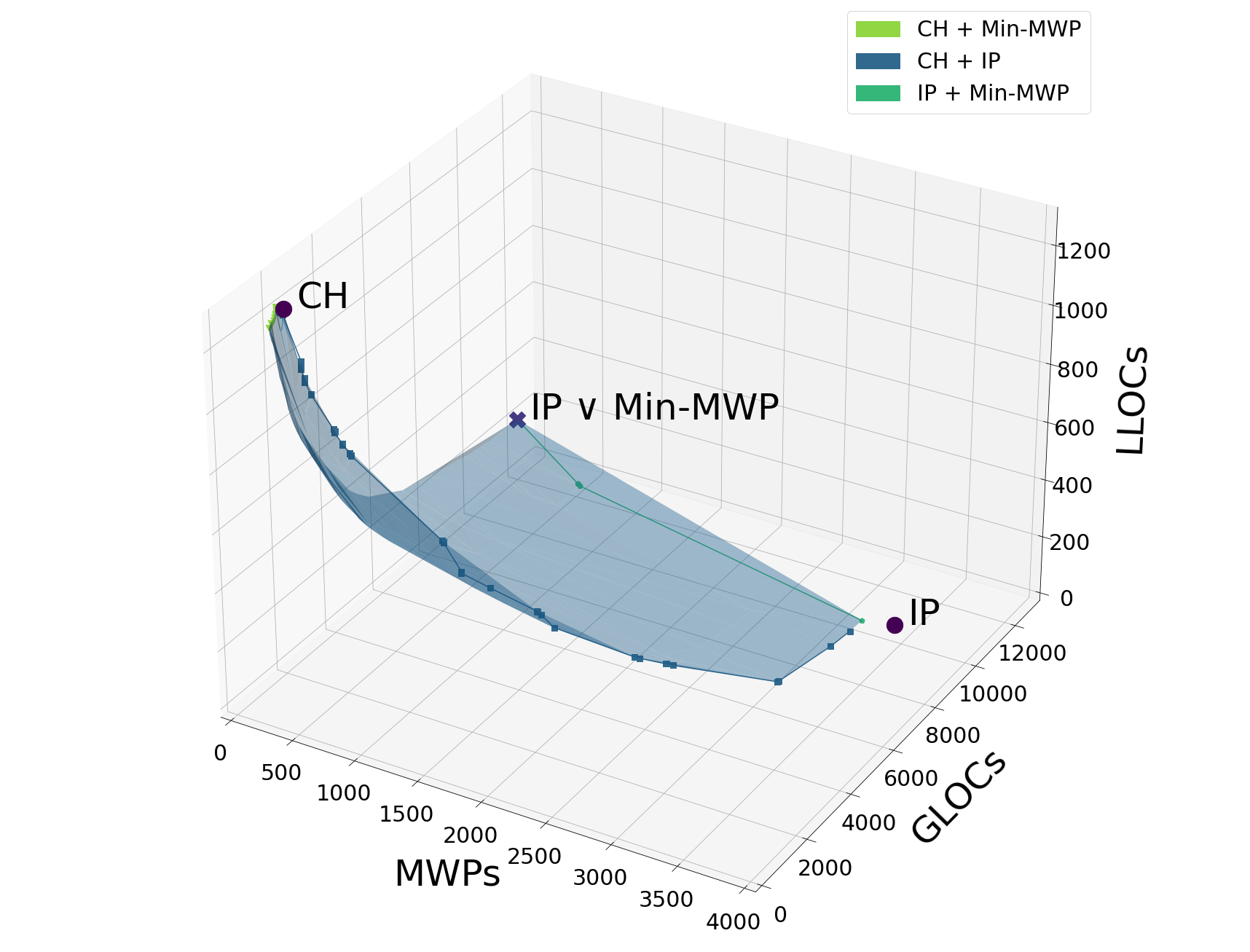}
		\caption{GLOCs, MWPs, LLOCs}
		\label{subfig:gloc_mwp_lloc}
	\end{subfigure}\hfill
	\begin{subfigure}{.475\linewidth}
		\includegraphics[width=\linewidth]{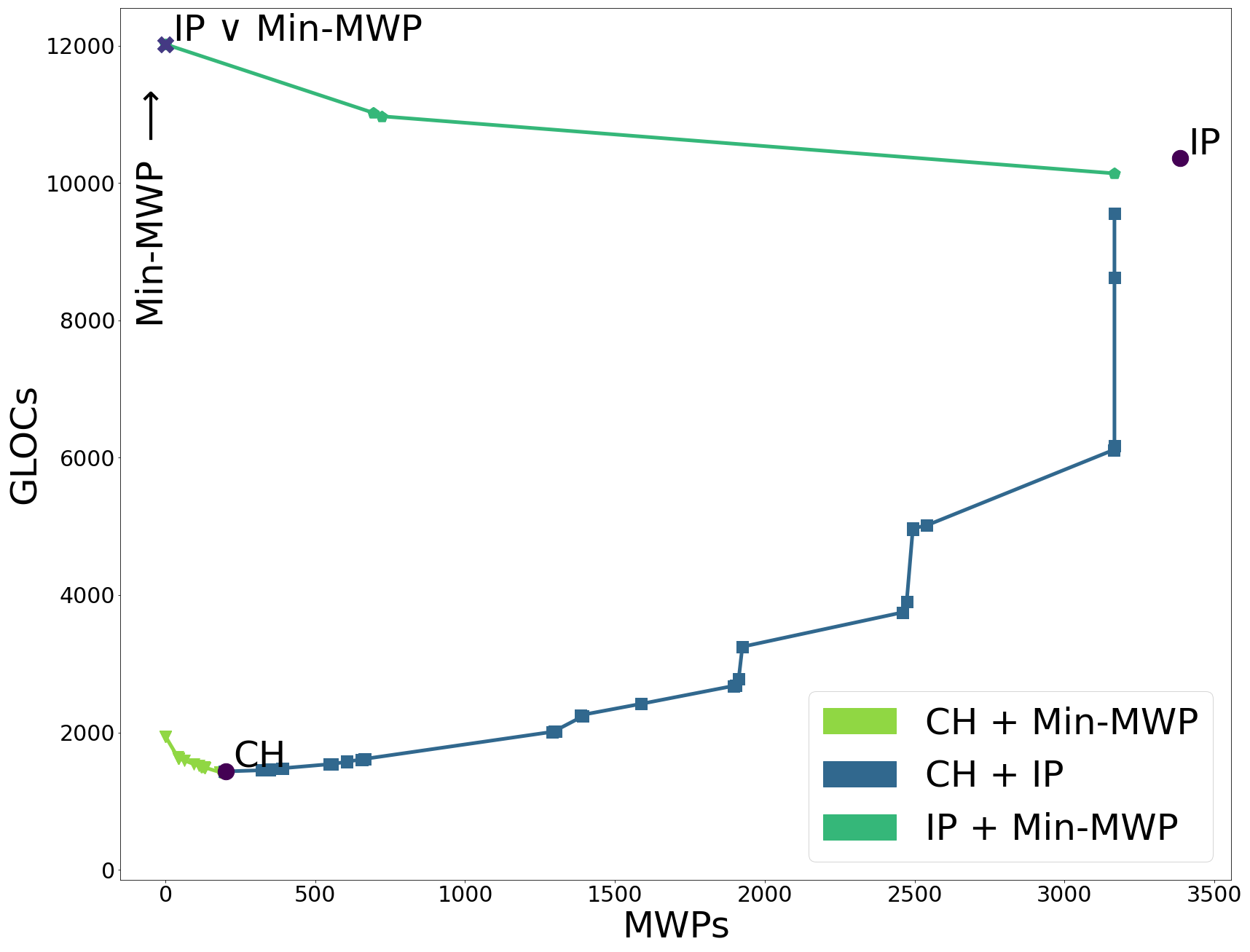}
		\caption{GLOCs, MWPs}
		\label{subfig:gloc_mwp}
	\end{subfigure}
	
	\medskip
	\begin{subfigure}{.475\linewidth}
		\includegraphics[width=\linewidth]{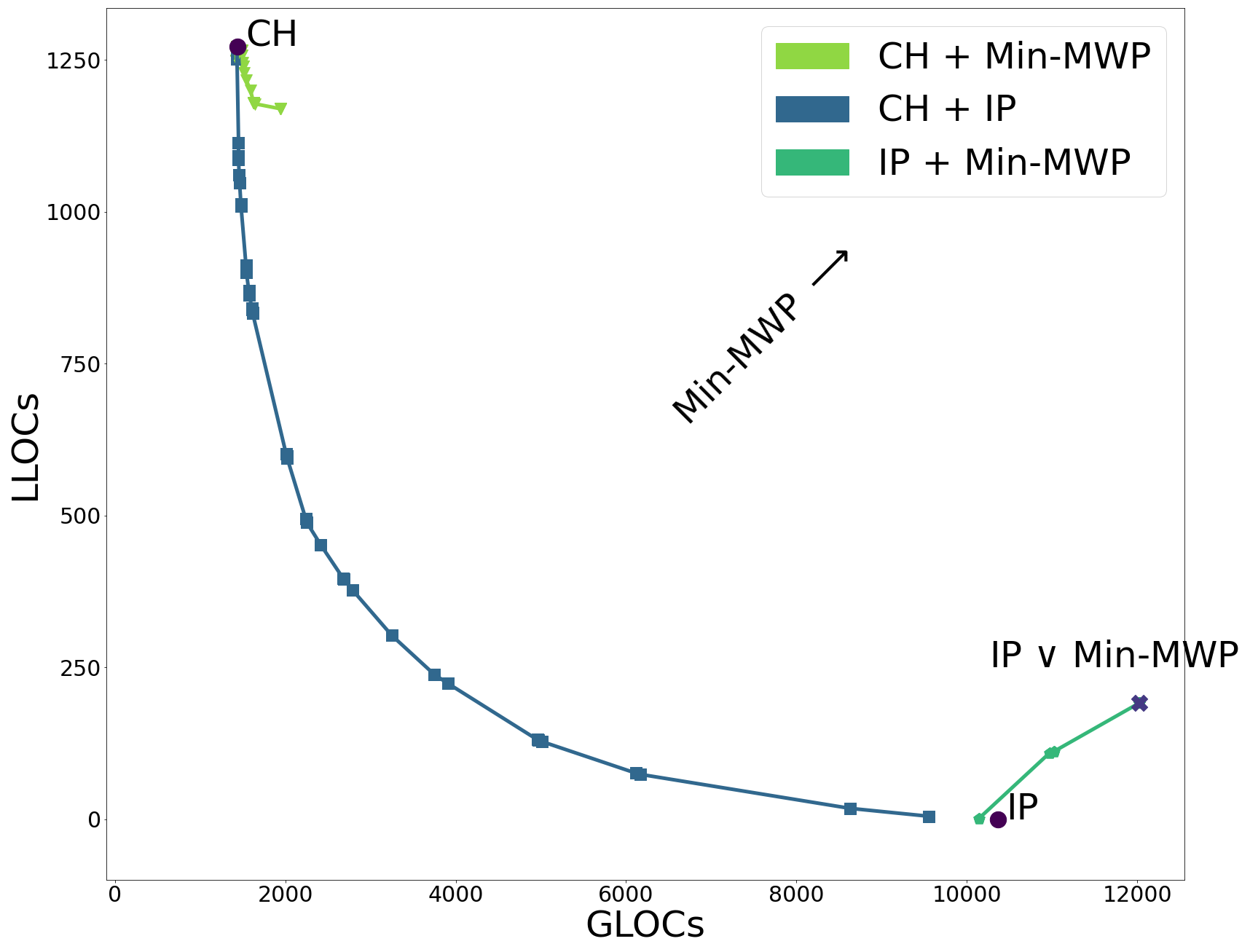}
		\caption{GLOCs, LLOCs}
		\label{subfig:gloc_lloc}
	\end{subfigure}\hfill
	\begin{subfigure}{.475\linewidth}
		\includegraphics[width=\linewidth]{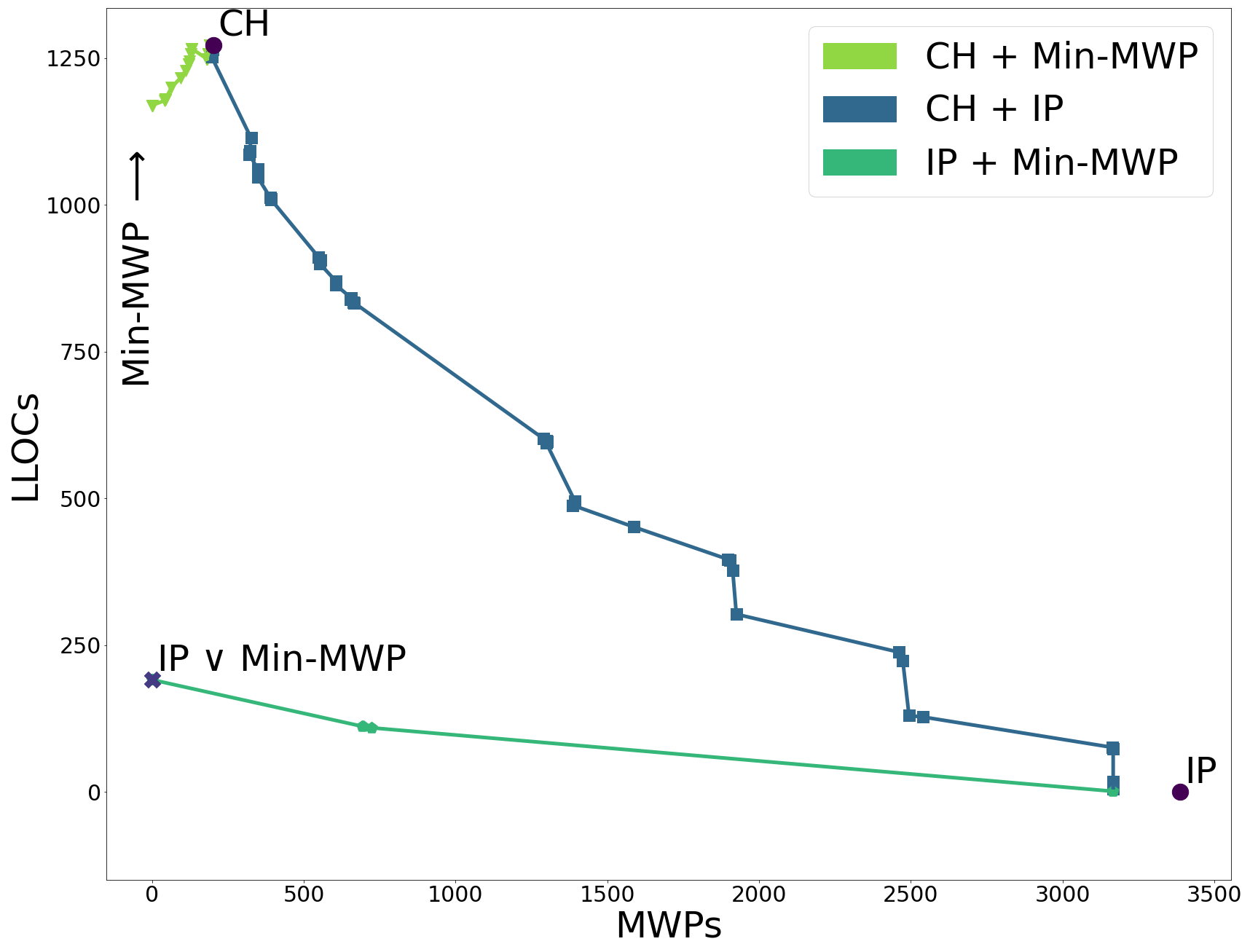}
		\caption{MWPs, LLOCs}
		\label{subfig:mwp_lloc}
	\end{subfigure}
	\caption{RTS System - GLOCs, MWPs, LLOCs [\$]}
	\label{fig:ieee_3d}
\end{figure}
} {}
 
For example, with an equal-weight linear scalarization of IP and CH pricing, we obtain prices that have less GLOCs and LLOCs than the average over IP and CH. This effect is particularly prominent for Min-MWP pricing: although the original price profile is fairly extreme, a linear scalarization causes the GLOCs and LLOCs to collapse very fast.\footnote{We also tested linear scalarizations with a non-exact ELMP approximation of CH pricing, in order to simulate a situation when CH prices are not readily available. This provided similar results, albeit with slightly increased GLOCs due to the non-exactness of the approximation.}

However, finding the correct weight vector for an a priori pricing rule requires information on preferences, which may require some degree of exploration and thus poses an impediment in practice. In contrast, the proposed join IP $\lor$ Min-MWP does not require such preference elicitation. 

\ifbool{figuresInText}{
\begin{figure}[H]	
	\centering
	\includegraphics[width=0.7\linewidth]{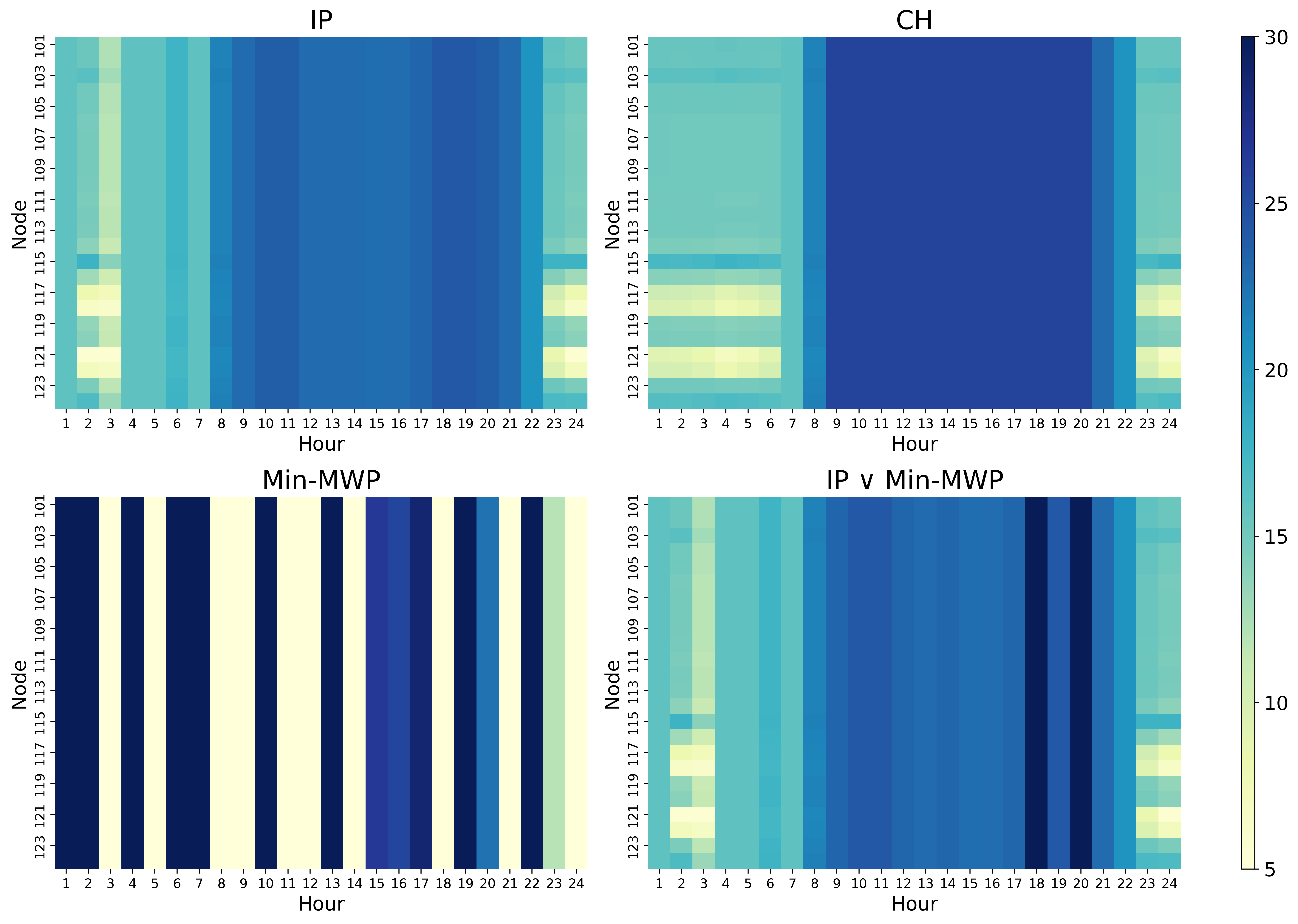}
	\caption{RTS System - Price Heatmaps [\$/MWh]}
	\label{fig:ieee_natural}
\end{figure}
}{}

Figure \ref{fig:ieee_natural} contains price heatmaps (in \$/MWh) along the different nodes and periods.
The Min-MWP pricing without additional constraints leads to high price volatility and inadequate congestion signals. However, joining it with standard IP pricing retains the congestion signals almost perfectly and at the same time reduces the substantial MWPs that IP pricing implies. The join further improves the congestion signals compared to CH pricing. For example, in hours 4-6 the single join price correctly signals that no congestion occurs in the network (as is also signaled by the single IP price), while CH pricing exhibits price differences and thus wrongly implies congestion.  

\subsection{European Bid Data} \label{subsec:bzr}

Next, we report results for a \ref{opt:DCOPFprimal} for the German bidding zone. The model is based on data published for the ongoing bidding zone review (BZR) in the European Union (EU). The BZR describes the process of evaluating the current bidding zone configuration as a result of structural congestion and efficiency losses in the EU day-ahead market. In this context, the European Network of Transmission System Operators for Electricity (ENTSO-E) conducted a nodal pricing study and published the related data in September 2022. \footnote{The data sets are available here: \url{https://www.entsoe.eu/network_codes/bzr/}.} We calibrate our \ref{opt:DCOPFprimal} model based on these data. 

The data sets contain hourly aggregated demand data as well as generator characteristics for 24 representative weeks of the years 1989, 1995, and 2009, as well as a network grid model for the German transmission system. We utilize these data to construct the \ref{opt:DCOPFprimal} as provided above. We obtain a system with 1687 nodes and 3232 transmission lines as well as bids for $24 \times 7 = 168$ days. 

We exemplarily report the result for three days: February 18, 2009, as a representative uncongested day according to the ENTSO-E study; July 30, 2009, as a representative congested day; and January 23, 2009 as a representative average day. The resulting MILPs of the \ref{opt:DCOPFprimal} are large with 4,538 generators, 1444 consumers, and more than 850,000 continuous and 200,000 binary variables for January 23, 2009. 
To reduce the computational effort\footnote{Note that in our experiments, the computational effort of the six-hour MILPs was negligible with runtime of only a few minutes on a standard computer, yet the memory requirements of computing all lost opportunity costs were high.} and memory requirements, we therefore split the 24-hour interval into four consecutive 6-hour intervals. In particular, instead of the regular 24-hour time horizon in European day-ahead markets, we use six-hour intervals and aggregate the respective lost opportunity costs. Technically, this means that intertemporal constraints are not enforced between the different six-hour intervals, but the impact on total lost opportunity costs is minor. Tables \ref{tab:bzr_stats_Feb18}, \ref{tab:bzr_stats_Jul30}, and \ref{tab:bzr_stats_Jan23} provide an overview of the results.

\ifbool{tablesInText}{
	\begin{table}[!htp]
		\centering
		\begin{tabular}{l|lll}
			& GLOCs [\$] & MWPs [\$] & LLOCs [\$]\\ \hline
			CH & 124434.83 & 57787.65 & 97296.52 \\
			IP & 432800.35 & 3026.59 & 0.00 \\ 
			Min-MWP & 4702804.83 & 0.00 & 7859610.21 \\
			\hline
			0.9 IP + 0.1 Min-MWP & 568788.64 & 1008.27 & 263.43 \\
			IP $\lor$ Min-MWP & 461757.95 & 653.33 & 732.44			
		\end{tabular}
		\caption{BZR Data - February 18, 2009 - GLOCs, MWPs, LLOCs}
		\label{tab:bzr_stats_Feb18}
	\end{table}
} {}

\ifbool{tablesInText}{
	\begin{table}[!htp]
		\centering
		\begin{tabular}{l|lll}
			& GLOCs [\$] & MWPs [\$] & LLOCs [\$]\\ \hline
			CH & 68392.90 & 2735.06 & 33562.90 \\
			IP & 184097.08 & 3982.19 & 0.00 \\ 
			Min-MWP & 6736749.48 & 0.00 & 9955661.19 \\
			\hline
			0.9 IP + 0.1 Min-MWP & 233327.03 & 854.46 & 128.50 \\
			IP $\lor$ Min-MWP & 220355.71 & 200.27 & 230.73
		\end{tabular}
		\caption{BZR Data - July 30, 2009 - GLOCs, MWPs, LLOCs}
		\label{tab:bzr_stats_Jul30}
	\end{table}
} {}

\ifbool{tablesInText}{
	\begin{table}[!htp]
		\centering
		\begin{tabular}{l|lll}
			& GLOCs [\$] & MWPs [\$] & LLOCs [\$]\\ \hline
			CH & 79158.72 & 24577.01 & 44726.50 \\
			IP & 514294.88 & 22487.21 &  0.00  \\ 
			Min-MWP & 11166195.10 & 0.00 & 8933860.68 \\
			\hline
			0.9 IP + 0.1 Min-MWP & 532385.17 & 2907.31 & 541.85 \\
			IP $\lor$ Min-MWP & 523313.00 & 325.51 & 1292.38 
		\end{tabular}
		\caption{BZR Data - January 23, 2009 - GLOCs, MWPs, LLOCs}
		\label{tab:bzr_stats_Jan23}
	\end{table}
} {}

The results confirm what was found for the IEEE RTS system. Due to the large size of this test system, the trade-offs between GLOCs, MWPs, and LLOCs are even more evident. In particular, as demand is price-inelastic, Min-MWP provides a solution with zero MWPs \citep{Bichler.2021}. 
The GLOCs and LLOCs implied by Min-MWP pricing are very high, yet adding a small weight of make-whole payments to IP can already reduce MWPs significantly. 
IP prices can lead to large MWPs, which are reduced by linear scalarization and the join, leading to only a small increase in LLOCs. 
As discussed earlier, linear scalarization is sensitive to the weights which are difficult to set a priori. 
In contrast, IP $\lor$ Min-MWP requires no parameterization and trades off MWPs and congestion signals (by means of LLOCs) in a meaningful way. The join also reduces GLOCs compared to IP and Min-MWP. 

\begin{result}
		Pure minimization of MWPs leads to very high GLOCs and LLOCs. Market participants have high incentives to deviate from the optimal outcome and congestion signals are flawed. The join IP $\lor$ Min-MWP significantly reduces LLOCs compared to Min-MWP pricing and MWPs compared to IP pricing. It produces a better congestion signal than CH pricing, while exhibiting lower MWPs at the same time. The differences in the lost opportunity costs between the pricing rules are more pronounced for the large instances of the German bidding zone compared to the small IEEE RTS data set.
\end{result}

Compared to IP $\lor$ Min-MWP, CH prices compromise on MWPs and LLOCs (and thus congestion signals) to reduce GLOCs. Typically, CH prices are intractable. As indicated earlier, we chose a model formulation that allows to compute CH prices in polynomial time. For more complex valuation or cost functions, convex hull formulations are not readily available, rendering CH pricing intractable \citep{Schiro.2016}. In contrast, IP $\lor$ Min-MWP can always be computed in polynomial time and it is practically tractable even for large problem sizes. 
\section{Conclusions} \label{sec:Conclusion}

The welfare theorems provide a solid foundation for the analysis of electricity market prices as they introduce the design desiderata one can hope for with a market-based allocation. 
A version of the welfare theorems for coupled markets delineates those environments for which we can expect Walrasian equilibria to exist from those where we do not. These theorems neither need linearity nor differentiability, but convexity of the cost and value functions. In particular, Walrasian prices cease to exist in the presence of non-convexities. Electricity spot markets are coupled and non-convex, and consequently there is no linear price vector that would be budget-balanced and envy-free. 
With linear and anonymous market prices on non-convex power markets, some participants require an uplift to compensate their loss, which incurs a budget deficit for the seller. 
In addition, the congestion signals on lines are biased, which results in non-zero LLOCs. This means, prices between nodes might differ even if no congestion occurs and the price difference does not reflect the marginal value of transmission capacity. 

The extended welfare theorems provide a foundation to classify different pricing rules for electricity markets as they have been suggested in the literature. We show that existing pricing rules minimize certain classes of lost opportunity costs, which can cause substantial increases in other relevant lost opportunity costs. 
Based on this trade-off, we propose to view the design of pricing rules in non-convex electricity markets as a multi-objective optimization problem. 

We analyze linear scalarization as a weighted sum of individual objective functions to derive a Pareto frontier of the different design goals. As this comes with certain practical limitations, i.e. the need for preference elicitation, we propose a novel and scalable pricing rule that requires no parametrization or fine-tuning. This join of the IP and Min-MWP pricing rule addresses current policy issues in U.S. and European electricity markets. In particular, the join minimizes side-payments, maintains adequate congestion signals, and can be computed efficiently. Based on realistic data sets, we demonstrate the possibilities of the join to capitalize on the upsides of several pricing rules. In view of recent concerns by regulators regarding increasing levels of MWPs with the IP pricing rule and the desire to maintain good congestion signals, the join provides a straightforward and easy-to-implement alternative for regulators. 

\subsection*{Acknowledgments}

The financial support from the German Research Foundation (Deutsche Forschungsgemeinschaft, DFG) (BI 1057/9-1) is gratefully acknowledged.

\bibliographystyle{informs2014} 
\bibliography{bibliography}

\appendix

\section{Economic Interpretability Problem} \label{app:economic_interpretability}

In Section \ref{sec:combinations}, we had mentioned that a dual pricing problem induces a convexified model for the underlying welfare maximization problem (\ref{opt:WM}), and that a poor choice of multi-objective formulation for the dual pricing may expand the feasible region of the underlying welfare maximization problem, exaggerating the gap between the dual pricing and the primal allocation problems. This, in turn, may lead to a loss of \emph{semantic meaning}, disallowing any meaningful economic interpretation. In this section of the electronic companion, we elaborate on this phenomenon. In order to do so, we first explain how to derive the corresponding convex model of a dual pricing problem by application of results from Section \ref{sec:Prelim}. We then illustrate how severe distortions in the welfare maximization problem may occur by considering two commonly used multi-objective solution concepts.

\subsection{Derivation of convex market models}\label{app:convex-models}

To derive the welfare maximization problem corresponding to some dual pricing problem $\min_{p} \sum_{\ell \in L} \lambda_\ell(p|z^*_\ell)$, we note that for a convex market the dual pricing functions are defined as in (\ref{def:dual-pricing}). Using convex conjugates, we can rewrite this as follows: 
\begin{align*}
	\lambda_b(p|x^*_b) & = (-v_b)^*(-p) + p^T x^*_b - v_b(x^*_b) & & \ \forall \ b \in B, \\
	\lambda_s(p|y^*_s) & = c^*_s(p) + c_s(y^*_s) - p^T y^*_s & & \ \forall \ s \in S, \\
	\lambda_r(p|f^*_r) & = d^*_r(B^T_r p) + d_r(f^*_r) - p^T B_r f^*_r & & \ \forall \ r \in R.
\end{align*}
This implies that, given a convex dual pricing function $\tilde{\lambda}_b(p|x^*_b)$ for a buyer $b$, denoting by $\tilde{v}_b$ the corresponding concave valuation function for buyer $b$,
\newcommand{\tv}{\tilde{v}}
\newcommand{\tc}{\tilde{c}}
\newcommand{\td}{\tilde{d}}
\newcommand{\tl}{\tilde{\lambda}}
\begin{equation}
	(-\tv_b)^*(p) = \tl_b(-p|x^*_b) + p^T x^*_b + v_b(x^*_b),  
\end{equation}
where $v_b(x^*_b)$ is the value buyer $b$ has for allocation $x^*_b$ in the \emph{original} welfare maximization problem (\ref{opt:WM}). Noting that for a closed convex function $f$ the biconjugate equals the function itself, i.e. $f = f^{**}$, we conclude that
\begin{align}
	\tv_b(x) & = -(\tl_b(-p|x^*_b) + p^T x^*_b + v_b(x^*_b))^*(x) \label{def:conv-buy} \\
	& = -(\max_{p} p^T (x - x^*_b) - v_b(x^*_b) - \tl_b(-p|x^*_b)) \nonumber \\
	& = -\tl^*_b(x^*_b - x) + v_b(x^*_b). \nonumber
\end{align} 
By a similar analysis, the corresponding cost function for a seller $s$ is given by 
\begin{equation}
	\tc_s(y) = \tl^*_c(y - y^*_s) + c_s(y^*_s).
\end{equation}
The induced cost functions of transmission operators, in turn, need to be more carefully defined due to the presence of matrices $B_r$ which specify the interaction of flows with the supply-demand constraints. In this case, as a first attempt one may consider a modified conjugate function
$$ \td_r(f) = \max_p p^T B_r (f - f^*_r) + d_r(f^*_r) - \tl_r(p|f^*_r).$$
However, this allows for transmission operator $r$ a feasible flow of the form $f = f' + \delta$, where $B_r \delta = 0$ (i.e. $\delta$ is in the \emph{null-space} $\textnormal{null}(B_r)$ of $B_r$) and $f'$ is orthogonal to $\delta$. Such flows $\delta$ do not have any effect on the supply-demand balance, and can be considered to be infeasible. Therefore, to rectify this, we instead set 
\begin{equation}
	\td_r(f) = \begin{cases}
		\max_p p^T B_r (f - f^*_r) + d_r(f^*_r) - \tl_r(p|f^*_r) & \textnormal{if } f^T \delta = 0, \forall \ \delta \in \textnormal{null}(B_r), \\
		+\infty & \textnormal{else.}
	\end{cases}
\end{equation}
\newcommand{\tom}{\tilde{\omega}}

Then by Proposition \ref{prop:calculus}.4, we note for the welfare function
\begin{align*}
	\tom(\sigma) = \max_{x,y,f} & \ \sum_{b \in B} \tv_b(x_b) - \sum_{s \in S} \tc_s(y_s) - \sum_{r \in R} \td_r(f_r) \\
	\text{ subject to} & \ \sum_{s \in S} y_s - \sum_{b \in B} x_b + \sum_{r \in R} B_r f_r = \sigma,
	\end{align*}
the conjugate of the negative welfare function $(-\tom)^*$ equals
\begin{equation}
	(-\tom)^*(p) = \sum_{\ell \in L} \tl_\ell(p|z^*_\ell) + \omega(0).
\end{equation}
The pricing problem for this convex market, given by the subgradient problem (\ref{opt:LDWM}), is therefore precisely $\min_{p} \sum_{\ell \in L} \tl_\ell(p|z^*_\ell)$.

To illustrate how the calculations work, we derive the convex model corresponding to (\ref{opt:min-MWP}):

\begin{example}
	In (\ref{opt:min-MWP}), each market participant $\ell$ has a dual pricing function $\lambda_\ell^{MWP}(p|z^*_\ell) = \max\{-u_\ell(z^*_\ell|p) , 0  \}$. For any buyer $b$, 
	$$u_b(x^*_b) = v_b(x^*_b) - p^T x^*_b.$$
	We know that the convex model $\tv_b$ here satisfies (\ref{def:conv-buy}), and by Proposition \ref{prop:calculus}.5, $[\lambda_b^{MWP}(\cdot)]^*(x) = \ch \min \{ (-u_b(x^*_b|\cdot))^*, 0^* \}(x)$. The convex conjugate of the identical zero function is $\chi_{\{0\}}(x)$, the indicator function for the singleton set containing $0$. Meanwhile, $(-u_b(x^*_b|\cdot))^*$ is given by 
	\begin{align*}
		(-u_b(x^*_b|\cdot))^*(x) & = \max_{p} p^T x + u_b(x^*_b|p) \\
		& = \max_p p^T (x - x^*_b) + v_b(x^*_b) \\
		& = \chi_{\{x^*_b\}}(x) + v_b(x^*_b).
	\end{align*}
	Therefore, substituting for the expression (\ref{def:conv-buy}), we get 
	$$-\tv_b(x) = \ch \min \{ \chi_{\{0\}}(x) + v_b(x^*_b) , \chi_{\{x^*_b\}} \} - v_b(x^*_b),$$
	which then implies that 
	$$ \tv_b(x) = \conch \max \{ -\chi_{\{0\}} , -\chi_{\{x^*_b\}} + v_b(x^*_b) \}.$$
	Specifically, buyer $b$'s valuations are modelled as the convexification of the valuation they have for their allocation, and the valuation they have for not participating in the market. Likewise, for a seller $s$ and a transmission operator $r$, we have
	\begin{align*}
		\tc_s(y) & = \ch \min \{ \chi_{\{0\}} , \chi_{\{y^*_s\}} + c_s(y^*_s) \}, \\
		\td_r(f) & = \ch \min \{ \chi_{\{0\}} , \chi_{\{f^*_r\}} + d_r(f^*_r) \}.
	\end{align*}
\end{example}

Thus it is indeed possible to extract a convex welfare maximization problem from a given dual pricing problem, and the dual pricing problem prices the optimal outcome of this welfare maximization problem. The set of participants and goods of this welfare maximization problem might in fact greatly differ for this problem; we provide two examples.

\begin{example}[Penalty functions: Fictitious goods and agents]
	As mentioned previously, in non-convex markets a Walrasian equilibrium need not exist. In this case, there is no price vector $p$ such that the conditions of Walrasian equilibria (\ref{def:WE}) are all satisfied. While we consider budget balance in linear payments, market participants' losses still need to be compensated. To finance these uplift payments, \citet{o2016dual} consider imposing personalized price vectors to participants. Specifically, their approach involves finding a set of participants who require make-whole payments, and then imposing personalized prices so that their uplifts are financed via the payments of other market participants. As a measure of fairness, they minimize the magnitude of these transfers. 

	Here, we analyze the distortionary effect of a similar pricing problem, stylized for simplicity of analysis. We impose personalized prices to the dual pricing problem (\ref{opt:min-GLOC}), with a quadratic penalty term for the magnitude of differences in prices. Furthermore, we add a budget balance constraint. Then the resulting dual pricing problem is given by 
	\begin{align*}
		\min_{(p_\ell)_{\ell \in L}} & \ \sum_{\ell \in L} \lambda^{CH}_\ell(p_\ell | z^*_\ell) + \frac{1}{2} \sum_{(\ell, \ell') \in L^2} \| p_\ell - p_{\ell'} \|^2_2 \\
		\textnormal{subject to} & \ \sum_{\ell \in L} p^T_\ell z^*_\ell = 0.
	\end{align*}
	We can then compute the associated welfare maximization problem for this dual pricing problem by following the procedure in Section \ref{app:convex-models}. The corresponding market has set of \emph{personalized} goods and flow parameters $M \times L$, $F \times L$. We also add a fictitious exchange operator $(\ell,\ell') \in L^2$ for each pair of participants, who can exchange goods and flows in $\{\ell\}\times (M \cup F)$ with the corresponding goods and flows in $\{\ell'\} \times (M \cup F)$ at a one-to-one ratio. Finally, we add an auctioneer $A$ who can provide any multiple of $(z^*_\ell)_{\ell \in L}$ to the market at cost $0$, clearing personalized goods in proportion to the priced optimal outcome. The resulting welfare maximization problem is thus
	\begin{align*}
		\max_{x,y,f, z_A, \large(z_{(\ell,\ell')}\large)_{(\ell,\ell') \in L^2}} & & \ \sum_{b \in B} v_b(x_b) - \sum_{s \in S} c_s(y_s) - \sum_{r \in R} d_r(f_r) & - \frac{1}{2} \sum_{(\ell,\ell') \in L^2} \|z_{(\ell,\ell')}\|^2_2 \\
		\text{subject to} & & \ x_b - z_A x^*_b - \sum_{\ell \in L} z_{(b,\ell)} - z_{(\ell,b)} & = 0 \ \forall \ b \in B \\ 
		& & \ y_s - z_A y^*_s - \sum_{\ell \in L} z_{(s,\ell)} - z_{(\ell,s)} & = 0 \ \forall \ s \in S \\ 
		& & \ f_r - z_A f^*_r - \sum_{\ell \in L} z_{(r,\ell)} - z_{(\ell,r)} & = 0 \ \forall \ r \in R. 
	\end{align*}
	The additional exchange operators may be thought of as participants who can take advantage of personalized price differences for arbitrage, while the auctioneer attempts to enforce optimal market clearing.
\end{example}

\begin{example}[Chebyshev scalarization: Consolidation of agents]
	To obtain a more balanced outcome with respect to the GLOCs of market participants, one might consider minimizing the (weighted) maximum of participants' GLOCs instead. Such a method is known as \emph{Chebyshev scalarization} in the literature, where for objectives to $f_1, f_2, ..., f_n$ to be jointly minimized, one seeks a solution of 
	\begin{align*}
		\min_{q \in \mathcal{F}} \max_{1 \leq i \leq n} \frac{f_i(q)}{w_i}.
	\end{align*}
	Here, $(w_i)_{1 \leq i \leq n}$ are the weights on each objective $f_i$, signifying their relative importance, and $\mathcal{F}$ is the feasible region of the problem.

	Let us consider implementing Chebyshev scalarization as a dual pricing problem, where each participant's GLOCs are scaled by the $2$-norm of the participant's allocation. Intuitively, this minimizes the GLOCs incurred for each unit of good purchased. This leads us to consider the dual pricing problem
	\begin{align*}
		\min_p \max_{\ell \in L} \frac{\lambda^{CH}_\ell(p|z^*_\ell)}{\|z^*_\ell\|_2}.
	\end{align*}
	However, this dual pricing problem is not additively separable over the market participants. Assuming the market does not have transmission operators for simplicity and by evaluating the convex conjugate of the dual pricing problem equals 
	$$\conch \max \left\{ \frac{v_b(\|x^*_b\|_2 \cdot (\sigma + x^*_b) ) - v_b(x^*_b)}{\|x^*_b\|_2}  \right\}_{b \in B} \cup \left\{ -\frac{c_s(\|y^*_s\|_2 \cdot (\sigma + y^*_s)) - c_s(y^*_s)}{\|y^*_s\|_2}  \right\}_{s \in S}.$$
	It does not seem possible, however, to transform this expression into a utilitarian welfare function for the original set of market participants. In fact, the expression does not appear to admit a straightforward interpretation as any meaningful welfare function.  
\end{example}

By the examples above, we see that the addition of constraints and penalty functions lead to an addition of new agents in the primal welfare maximization problem, while combining participants' lost opportunity costs under a single term consolidates them into a single entity. We expect such drastic changes in the welfare maximization problem to widen the gap between the true optimal allocation and the allocation priced by the dual pricing problem, as in the case of CH pricing. Therefore, we would like to avoid such distortions if possible.

We note that the dimension $M$ of the price vector $p \in \mathbb{R}^M$ is precisely the number of priced goods. This is because the welfare function $\omega$ takes as its argument supply-demand constraint violations in real goods, and prices are in the corresponding dual space. To maintain the number of goods, we must thus impose linear prices -- as modifying the size of the price vector (e.g. by imposing personalized prices) modifies the number of goods in the corresponding welfare maximization problem.

To maintain the number of participants, note that by Proposition \ref{prop:calculus}.4 a welfare maximization problem with a set of agents $L$ necessarily has an associated subgradient problem (\ref{opt:LDWM}) that is additively separable over the participants. That is, for a welfare maximization problem of the form 
\begin{align*}
	\max_{x,y,f} & \ \sum_{b \in B} \tv_b(x_b) - \sum_{s \in S} \tc_s(y_s) - \sum_{r \in R} \td_r(f_r) \\
	\text{ subject to} & \ \sum_{s \in S} y_s - \sum_{b \in B} x_b + \sum_{r \in R} B_r f_r = 0,
\end{align*}
the dual pricing problem is of the form $\min_p \sum_{\ell \in L} \tl_\ell(p|z^*_\ell)$. The converse implication also holds and an additively separable dual pricing function leads to a welfare maximization problem with set of participants $L$, as shown by our discussion in \ref{app:convex-models}. This leads us to observe that, to obtain a dual pricing problem with minimal distortion, we should restrict attention to dual pricing problems with an additively separable objective function (over $L$) and linear prices for each good.

\section{DCOPF Problem} \label{app:DCOPF_Notation}

\begin{table}[H]
	\centering
	\begin{tabular}{lll}
		\hline
		Sets & & \\ 
		\hline
		$B$ & & Buyers \\
		$S$ & & Sellers \\
		$T = \{1,..,\overline{T}\}$ & & Time Periods \\
		$V$ & & Nodes (with some $R^* \in V$ as reference node) \\
		$N(v)$ & & Neighboring nodes of a node $v \in V$ \\
		$\beta_b^t$ & & Bids of buyer $b \in B$ in period $t \in T$ \\
		$\beta_s^t$ & & Bids of seller $s \in S$ in period $t \in T$ \\
		\hline
		Mappings & & \\
		\hline
		$\nu(b)$ & $B \rightarrow V$ & Mapping of buyer $b \in B$ to its node $v \in V$ \\
		$\nu(s)$ & $S \rightarrow V$ & Mapping of seller $s \in S$ to its node $v \in V$ \\
		\hline
		Parameters & & \\
		\hline
		$B_{vw}$ & [pu] & Susceptance of the line connecting $v,w \in V$ \\
		$v_{bt\ell}$ & [\$/MWh] & Value of bid $\ell \in \beta_b^t$ of buyer $b \in B$ in period $t \in T$ \\
		$q_{bt\ell}$ & [MWh] & Maximum quantity of bid $\ell \in \beta_b^t$ of buyer $b \in B$ in period $t \in T$ \\
		$\underline{P}_{bt}$ & [MWh] & Price-inelastic demand of buyer $b \in B$ in period $t \in T$ \\
		$\overline{P}_{bt}$ & [MWh] & Maximum demand of buyer $b \in B$ in period $t \in T$ \\
		$c_{st\ell}$ & [\$/MWh] & Cost of bid $\ell \in \beta_s^t$ of seller $s \in S$ in period $t \in T$ \\
		$h_s$ & [\$] & No-load costs of seller $s \in S$ \\
		$q_{st\ell}$ & [MWh] & Maximum quantity of bid $\ell \in \beta_s^t$ of seller $s \in S$ in period $t \in T$ \\
		$\underline{P}_{st}$ & [MWh] & Minimum output of seller $s \in S$ in period $t \in T$ \\
		$\overline{P}_{st}$ & [MWh] & Maximum output of seller $s \in S$ in period $t \in T$ \\
		$\underline{R}_s$ & & Minimum uptime of seller $s \in S$ \\
		$\underline{F}_{vw}$ & [MWh] & Minimum flow on the line connecting $v,w \in V$ \\
		$\overline{F}_{vw}$ & [MWh] & Maximum flow on the line connecting $v,w \in V$ \\
		\hline
		Primal Variables & & \\
		\hline
		$x_{bt} \geq 0$ & [MWh] & Consumption of buyer $b \in B$ in period $t \in T$ \\
		$x_{bt\ell} \geq 0$ & [MWh] & Consumption of buyer $b \in B$ in period $t \in T$ regarding bid $\ell \in \beta_b^t$ \\
		$y_{st} \geq 0$ & [MWh] & Generation of seller $s \in S$ in period $t \in T$ \\
		$y_{st\ell} \geq 0$ & [MWh] & Generation of seller $s \in S$ in period $t \in T$ regarding bid $\ell \in \beta_s^t$ \\
		$u_{st} \in \{0,1\}$ & & Commitment of seller $s \in S$ in period $t \in T$ \\
		$\phi_{st} \geq 0$ & & Start-up indicator for seller $s \in S$ in period $t \in T$ \\
		$\alpha_{vt} \in \mathbb{R}$ & [rad] & Voltage angle at node $v \in V$ in period $t \in T$ \\
		$f_{vwt} \in \mathbb{R}$ & [MWh] & Flow on the line connecting $v,w \in V$ in period $t \in T$ \\
		\hline
		Dual Variables & & \\
		\hline
		$p_{vt} \in \mathbb{R}$ & [\$/MWh] & Price at node $v \in V$ in period $t \in T$ \\
		$\gamma_{vwt} \in \mathbb{R}$ & [\$/MWh] & Congestion price for the line connecting $v,w \in V$ in period $t \in T$ \\
		$r_t \in \mathbb{R}$ & & Dual of the reference node voltage angle constraint in period $t \in T$ 
	\end{tabular}
	\caption{DCOPF Notation}
	\label{tab:DCOPF_Notation}
\end{table}

\begin{align}
	\max_{} \quad & \sum_{b \in B} \sum_{t \in T} \sum_{\ell \in \beta^t_b} v_{bt\ell} x_{bt\ell} - \sum_{s \in S} \sum_{t \in T} \sum_{\ell \in \beta^t_s} c_{st\ell} y_{st\ell} - \sum_{s \in S} \sum_{t \in T} h_s u_{st} \label{opt:DCOPF_Primal} & \tag{DCOPF-MILP} \\
	\text{subject to \quad} & x_{bt\ell} \in [0,q_{bt\ell}] & \forall b \in B, t \in T, \ell \in \beta_b^t \nonumber \\
	& x_{bt} - \sum_{\ell \in \beta_b^t} x_{bt\ell} = \underline{P}_{bt} & \forall b \in B, t \in T \nonumber \\
	& x_{bt} \leq \overline{P}_{bt} & \forall b \in B, t \in T \nonumber \\
	& y_{st\ell} \geq 0 & \forall \ s \in S, t \in T, \ell \in \beta_s^t \nonumber \\
	& y_{st\ell}  - u_{st} q_{st\ell} \leq 0 & \forall \ s \in S, t \in T, \ell \in \beta_s^t \nonumber \\
	& y_{st} - \sum_{\ell \in \beta_s^t} y_{st\ell} = 0 & \forall \ s \in S, t \in T \nonumber \\
	& y_{st} - \underline{P}_{st} u_{st} \geq 0 & \forall \ s \in S, t \in T \nonumber \\
	& y_{st} - \overline{P}_{st} u_{st} \leq 0 & \forall \ s \in S, t \in T \nonumber \\
	& \phi_{st} - u_{st} + u_{s(t-1)} \geq 0 & \forall \ s \in S, 1 < t \leq T \nonumber \\
	& \sum_{i = t - R_s + 1}^t \phi_{si} - u_{st} \leq 0 & \forall \ s \in S, 1 < t \leq T \nonumber \\
	& f_{vwt} \in [\underline{F}_{vw}, \overline{F}_{vw}] & \forall \ v \in V, w \in N(v), t \in T \nonumber \\
	& f_{vwt} - B_{vw}(\alpha_{vt}-\alpha_{wt}) = 0 & \forall \ v \in V, w \in N(v), t \in T \nonumber \\
	& \sum_{s:\nu(s)=v} y_{st} - \sum_{b:\nu(b)=v} x_{bt} - \sum_{w \in N(v)} f_{vwt} = 0 & \forall \ v \in V, t \in T \nonumber \\
	& \alpha_{R^*t} = 0 & \forall \ t \in T \nonumber
\end{align}

\section{ELMP Pricing Problem} \label{app:ELMP-Dual}
The continuous relaxation of the binary integer constraints in the welfare maximization problem provide an LP. The associated dual pricing problem with this relaxation of the welfare maximization problem is the pricing rule known as ELMP \citep{MISO.2019}. Furthermore, the class of valuation / cost functions we consider are such that the optimal solutions to the dual LP provide CH prices \citep{Hua.2017}.

Therefore, to formulate the Convex Hull pricing problem in our setting we consider the dual LP to (\ref{opt:DCOPFprimal}) where for each seller $s$ and each time period $t$, the binary integer constraints are relaxed $u_{st} \in [0,1]$. Furthermore, we add constants to the dual objective function which do not alter the solution sets. However, the addition of these constants allows the objective function value to be sum of GLOCs. The resulting dual LP is given as:
\begin{align}
	\min_{\overline\epsilon, \overline\psi, \bar\chi \geq 0, \underline\epsilon, \underline\psi, \underline\chi,  \hat\chi \leq 0, p, \gamma, r, \lambda, \epsilon} \quad & \sum_{b \in B} \lambda_b + \sum_{s \in S} \lambda_s + \sum_{v \in V, w \in N(v), t \in T} \lambda_{vwt} \label{opt:DCOPFELMP} & \tag{ELMP-LP} \\
	\text{subject to \quad} &  \lambda_b - \sum_{t \in T} \Big[ \epsilon_{bt} \underline{P}_{bt} + \overline\epsilon_{bt} \overline{P}_{bt} + \sum_{\ell \in \nu_b^t} \overline\epsilon_{bt\ell}  q_{bt\ell}\Big] + v_b(x^*_b) - p^T_{\nu(b)} x^*_b \geq 0 \quad \quad \forall \ b \in B \nonumber \\
	& \lambda_s - \sum_{t \in T} \overline\psi_{st}  + p^T_{\nu(s)} y^*_s - c_s(y^*_s, u^*_s) \geq 0 \quad \quad \forall \ s \in S \nonumber \\
	& \lambda_{vwt} - \overline\epsilon_{vwt} \overline{F}_{vw} - \underline\epsilon_{vwt} \underline{F}_{vw} + \gamma_{vwt} f^*_{vwt} \geq 0 \quad \quad \forall \ v \in V, w \in N(v), t \in T \nonumber \\ 
	& \sum_{w | v \in N(w)} B_{wv} (p_{wt} + \gamma_{wvt}) - \sum_{w \in N(v)} B_{vw} (p_{vt} + \gamma_{vwt}) = 0 \quad \quad \forall \ v \in V \setminus \{R^*\}, t \in T \nonumber \\
	& r_t + \sum_{w | R^* \in N(w)} B_{wR^*} (p_{wt} + \gamma_{wR^*t}) - \sum_{w \in N(R^*)} B_{R^*w} (p_{R^*t} + \gamma_{R^*wt}) = 0 \quad \quad \forall \ t \in T \nonumber \\
	& -\gamma_{vwt} + \overline\epsilon_{vwt} + \underline\epsilon_{vwt} = 0 \quad \quad \forall \ v \in V, w \in N(v), t \in T \nonumber \\
	& \overline\epsilon_{bt\ell} + \underline\epsilon_{bt\ell} - \epsilon_{bt} = v_{bt\ell} \quad \quad \forall \ b \in B, t \in T, \ell \in \nu_b^t \nonumber \\
	& \epsilon_{bt} + \overline\epsilon_{bt} + p_{\nu(b)t} = 0 \quad \quad \forall \ b \in B, t \in T \nonumber \\
	& \overline\epsilon_{st\ell} + \underline\epsilon_{st\ell} - \epsilon_{st} = -c_{st\ell} \quad \quad \forall \ s \in S, t \in T, \ell \in \nu_s^t \nonumber \\
	& \epsilon_{st} + \underline\epsilon_{st} + \overline\epsilon_{st} - p_{\nu(s)t} = 0 \quad \quad \forall \ s \in S, t \in T \nonumber \\
	& -\sum_{\ell \in \nu_s^1} q_{s1\ell} \overline\varepsilon_{s1\ell} + \overline\psi_{s1} + \underline\psi_{s1} - \overline{P}_{s1} \overline\epsilon_{s1} - \underline{P}_{s1} \underline\epsilon_{s1} + \underline\chi_{s2} = -h_{s1}, \quad \quad \forall \ s \in S \nonumber \\
	& -\sum_{\ell \in \nu_s^t} q_{st\ell} \overline\varepsilon_{st\ell} + \overline\psi_{st} + \underline\psi_{st} - \overline{P}_{st} \overline\epsilon_{st} - \underline{P}_{st} \underline\epsilon_{st} - \overline\chi_{st} - \underline\chi_{st} + \underline\chi_{s(t+1)} = -h_{st} \nonumber \\
	& \quad \quad \quad \quad \quad \quad \quad \quad \quad\quad \quad \quad\quad \quad \quad \quad \quad \quad\quad \quad \quad \quad \quad \quad  \forall s \in S, 1 < t < T \nonumber \\
	& -\sum_{\ell \in \nu_s^T} q_{sT\ell} \overline\varepsilon_{sT\ell} + \overline\psi_{sT} + \underline\psi_{sT} - \overline{P}_{sT} \overline\epsilon_{sT} - \underline{P}_{sT} \underline\epsilon_{sT} - \overline\chi_{sT} - \underline\chi_{sT} = -h_{sT} \quad \quad \forall \ s \in S \nonumber \\
	& \hat\chi_{st} + \underline\chi_{st} + \sum_{t' = t}^{\min\{\overline{T},t + \underline{R}_s - 1\}} \bar\chi_{st} = 0 \quad \quad \forall \ s \in S, 1 < t \leq T. \nonumber
\end{align}

\section{IP Pricing Problem} \label{app:IP-Dual}
Note that in our setting, an optimal allocation $\left( (x^*_b)_{b \in B}, (y^*_s)_{s \in S}, (f^*_r)_{r \in R} \right)$ fixes the binary integer variables $u^*_{st}$ which indicate whether seller $s$ is generating at period $t$. IP pricing then restricts the primal welfare maximization problem (\ref{opt:DCOPFprimal}) by adding the constraints $u_{st} = u^*_{st} \ \forall \ s \in S, t \in T$. The resulting welfare maximization problem is an LP, and the dual LP provides prices which minimize LLOCs. Again, we modify the dual objective function to be the sum of LLOCs by adding constants that do not affect the optimal solution. In our setting, this dual LP is given as:
\begin{align}
	\min_{\overline\epsilon \geq 0, \underline\epsilon \leq 0, p, \gamma, r, \lambda, \epsilon} \quad & \sum_{b \in B} \lambda_b + \sum_{s \in S} \lambda_s + \sum_{v \in V, w \in N(v), t \in T} \lambda_{vwt} \label{opt:DCOPFIP3} & \tag{IP-LP} \\
	\text{subject to \quad} &  \lambda_b - \sum_{t \in T} \Big[ \epsilon_{bt} \underline{P}_{bt} + \overline\epsilon_{bt} \overline{P}_{bt} + \sum_{\ell \in \nu_b^t} \overline\epsilon_{bt\ell}  q_{bt\ell}\Big] + v_b(x^*_b) - p^T_{\nu(b)} x^*_b \geq 0 \quad \quad \forall \ b \in B \nonumber \\
	& \lambda_s - \sum_{t \in T} \Big[ \underline\epsilon_{st} \underline{P}_{st} u^*_{st} + \overline\epsilon_{st} \overline{P}_{st} u^*_{st} + \sum_{\ell \in \nu_b^t} \overline\epsilon_{st\ell} q_{st\ell} u^*_{st} \Big] + p^T_{\nu(s)} y^*_s - c_s(y^*_s, u^*_s) \geq - h^T u^*_s \quad \quad \forall \ s \in S \nonumber \\
	& \lambda_{vwt} - \overline\epsilon_{vwt} \overline{F}_{vw} - \underline\epsilon_{vwt} \underline{F}_{vw} + \gamma_{vwt} f^*_{vwt} \geq 0 \quad \quad \forall \ v \in V, w \in N(v), t \in T \nonumber \\ 
	& \sum_{w | v \in N(w)} B_{wv} (p_{wt} + \gamma_{wvt}) - \sum_{w \in N(v)} B_{vw} (p_{vt} + \gamma_{vwt})  = 0 \quad \quad \forall \ v \in V \setminus \{R^*\}, t \in T \nonumber \\
	& r_t + \sum_{w | R^* \in N(w)} B_{wR^*} (p_{wt} + \gamma_{wR^*t}) - \sum_{w \in N(R^*)} B_{R^*w} (p_{R^*t} + \gamma_{R^*wt}) = 0 \quad \quad \forall \ t \in T \nonumber \\
	& -\gamma_{vwt} + \overline\epsilon_{vwt} + \underline\epsilon_{vwt} = 0 \quad \quad \forall \ v \in V, w \in N(v), t \in T \nonumber \\
	& \overline\epsilon_{bt\ell} + \underline\epsilon_{bt\ell} - \epsilon_{bt} = v_{bt\ell} \quad \quad \forall \ b \in B, t \in T, \ell \in \nu_b^t \nonumber \\
	& \epsilon_{bt} + \overline\epsilon_{bt} + p_{\nu(b)t} = 0 \quad \quad \forall \ b \in B, t \in T \nonumber \\
	& \overline\epsilon_{st\ell} + \underline\epsilon_{st\ell} - \epsilon_{st} = -c_{st\ell} \quad \quad \forall \ s \in S, t \in T, \ell \in \nu_s^t \nonumber \\
	& \epsilon_{st} + \underline\epsilon_{st} + \overline\epsilon_{st} - p_{\nu(s)t} = 0 \quad \quad \forall \ s \in S, t \in T. \nonumber 
\end{align}

\section{Min-MWP Pricing Problem} \label{app:MWP-Dual}
As mentioned before, while a direct implementation of (\ref{opt:min-MWP}) is possible, solutions to the resulting dual pricing problem may result in phase angle operators having infinite LLOCs / GLOCs. To rectify this issue, in our implementation of (\ref{opt:min-MWP}) we still account for GLOCs of transmission operators with respect to phase angles. This is achieved by including the dual constraints associated with primal variables $\alpha_{vt}$ for $v \in V, t \in T$, and the resulting dual pricing problem is given as:
\begin{align}
	\min_{p, \gamma, r, \lambda} & & \sum_{b \in B} \lambda_b + \sum_{s \in S} \lambda_s + \sum_{v \in V, w \in N(v), t \in T} \lambda_{vwt} \label{opt:DCOPFPBE} \tag{Min-MWP-LP}& \\
	\text{subject to} & & -v_b(x^*_b) + p^T_{\nu(b)} x^*_b - \lambda_b & \leq 0 \ \forall \ b \in B \nonumber \\
	& & -p^T_{\nu(s)} y^*_s + c_s(y^*_s, u^*_s) - \lambda_s & \leq 0 \ \forall \ s \in S \nonumber \\
	& & -\gamma_{vwt} f^*_{vwt} - \lambda_{vwt} & \leq 0 \ \forall \ v \in V, w \in N(v), t \in T \nonumber \\
	& &\sum_{w | v \in N(w)} B_{wv} (p_{wt} + \gamma_{wvt}) - \sum_{w \in N(v)} B_{vw} (p_{vt} + \gamma_{vwt})  & = 0 \ \forall \ v \in V \setminus \{R^*\}, t \in T & \nonumber  \\
	& & r_t + \sum_{w | R^* \in N(w)} B_{wR^*} (p_{wt} + \gamma_{wR^*t}) - \sum_{w \in N(R^*)} B_{R^*w} (p_{R^*t} + \gamma_{R^*wt}) & = 0 \ \forall \ t \in T. & \nonumber
\end{align}

\section{IP $\lor$ Min-MWP Pricing Problem} \label{app:Join}
As discussed, the join of IP and Min-MWP pricing considers each participant's maximum of the IP and Min-MWP dual pricing function. For \ref{opt:DCOPFprimal}, we need not introduce new decision variables and the addition of only $|B \cup S|$ constraints to \ref{opt:DCOPFIP3} is sufficient. Then, by choice of the constraints, each participant's contribution to the objective function $\lambda_\ell$ will equal the maximum of their MWPs and LLOCs. The dual LP is given as: 

\begin{align}
	\min_{\overline\epsilon \geq 0, \underline\epsilon \leq 0, p, \gamma, r, \lambda, \epsilon} \quad & \sum_{b \in B} \lambda_b + \sum_{s \in S} \lambda_s + \sum_{v \in V, w \in N(v), t \in T} \lambda_{vwt} \label{opt:DCOPF_Join} & \tag{(IP $\lor$ Min-MWP)-LP} \\
	\text{subject to \quad} & \lambda_b - \sum_{t \in T} \Big[ \epsilon_{bt} \underline{P}_{bt} + \overline\epsilon_{bt} \overline{P}_{bt} + \sum_{\ell \in \nu_b^t} \overline\epsilon_{bt\ell}  q_{bt\ell}\Big] + v_b(x^*_b) - p^T_{\nu(b)} x^*_b \geq 0 \quad \quad \forall \ b \in B \nonumber \\
	& v_b(x^*_b) -  p^T_{\nu(b)} x^*_b + \lambda_b \geq 0 \quad \quad \forall \ b \in B \nonumber \\
	& \lambda_s - \sum_{t \in T} \Big[ \underline\epsilon_{st} \underline{P}_{st} u^*_{st} + \overline\epsilon_{st} \overline{P}_{st} u^*_{st} + \sum_{\ell \in \nu_b^t} \overline\epsilon_{st\ell} q_{st\ell} u^*_{st} \Big] + p^T_{\nu(s)} y^*_s - c_s(y^*_s, u^*_s) \geq - h^T u^*_s \quad \quad \forall \ s \in S \nonumber \\
	& p^T_{\nu(s)} y^*_s - c_s(y^*_s, u^*_s) + \lambda_s \geq 0 \quad \quad \forall \ s \in S \nonumber \\
	& \lambda_{vwt} - \overline\epsilon_{vwt} \overline{F}_{vw} - \underline\epsilon_{vwt} \underline{F}_{vw} + \gamma_{vwt} f^*_{vwt} \geq 0 \quad \quad \forall \ v \in V, w \in N(v), t \in T \nonumber \\ 
	& \sum_{w | v \in N(w)} B_{wv} (p_{wt} + \gamma_{wvt}) - \sum_{w \in N(v)} B_{vw} (p_{vt} + \gamma_{vwt}) = 0 \quad \quad \forall v \in V \setminus \{R^*\}, d t \in T \nonumber \\
	& r_t + \sum_{w | R^* \in N(w)} B_{wR^*} (p_{wt} + \gamma_{wR^*t}) - \sum_{w \in N(R^*)} B_{R^*w} (p_{R^*t} + \gamma_{R^*wt}) = 0 \quad \quad \forall \ t \in T \nonumber \\
	& -\gamma_{vwt} + \overline\epsilon_{vwt} + \underline\epsilon_{vwt} = 0 \quad \quad \forall \ v \in V, w \in N(v), t \in T \nonumber \\
	& \overline\epsilon_{bt\ell} + \underline\epsilon_{bt\ell} - \epsilon_{bt} = v_{bt\ell} \quad \quad \forall \ b \in B, t \in T, \ell \in \nu_b^t \nonumber \\
	& \epsilon_{bt} + \overline\epsilon_{bt} + p_{\nu(b)t} = 0 \quad \quad \forall \ b \in B, t \in T \nonumber \\
	& \overline\epsilon_{st\ell} + \underline\epsilon_{st\ell} - \epsilon_{st} = -c_{st\ell} \quad \quad \forall \ s \in S, t \in T, \ell \in \nu_s^t \nonumber \\
	& \epsilon_{st} + \underline\epsilon_{st} + \overline\epsilon_{st} - p_{\nu(s)t} = 0 \quad \quad \forall \ s \in S, t \in T. \nonumber 
\end{align}

\section{ARPA-E Grid Optimization Competition Data} \label{app:arpa}

Apart from the IEEE RTS System and the BZR data, we test our pricing rules on the network and bid data provided for the ARPA-E Grid Optimization Competition Challenge 2.\footnote{See \url{https://gocompetition.energy.gov/challenges/challenge-2} for further details.} This competition seeks the development of modern and scalable optimization techniques for solving complex power flow problems. To that end, they provide large-scale and realistic test sets of single-period power flow problems. For our purposes, we test the five different scenarios provided for an exemplary 617-node grid. We parameterize \ref{opt:DCOPFprimal} with the available data (with capped susceptance values to avoid numerical instability) and report the results of exemplary linear scalarizations and the join in the table below.  

\ifbool{tablesInText}{
	\begin{table}[H]
		\centering
		\resizebox{0.95\textwidth}{!}{
			\begin{tabular}{l|ccc|ccc|ccc|ccc|ccc}
				& & 617-I & & & 617-II & & \\
				& GLOCs & MWPs & LLOCs & GLOCs & MWPs & LLOCs  \\ \hline
				CH & 1393.93 & 856.66 & 1145.55 & 1878.76 & 1364.99 & 1273.89 \\
				IP &  7690.75 & 4913.05 & 0.0 & 6826.01 & 3783.67 & 0.0 \\ 
				Min-MWP & 830053.91 & 0.0 & 506335.11 & 611089.04 & 0.0 & 441164.65 \\
				\hline
				0.9 IP + 0.1 Min-MWP & 6029.08 & 3219.04 & 38.77 & 6275.68 & 3038.20 & 53.77 \\
				0.9 CH + 0.1 Min-MWP & 1393.97 & 852.46 & 1145.59 & 1879.28 & 1359.99 & 1274.33 \\
				0.5 IP + 0.5 CH & 1532.48 & 1169.44 & 871.31 & 2174.56 & 1722.28 & 819.39 \\ \hline
				IP $\lor$ Min-MWP & 3990.84 & 749.79 & 1093.35 & 5055.31 & 1053.66 & 1222.65
		\end{tabular}}
		\vspace{0.5cm}
		\resizebox{\textwidth}{!}{
			\begin{tabular}{l|ccc|ccc|ccc}
				& & 617-III & & & 617-IV & & & 617-V  \\
				& GLOCs & MWPs & LLOCs & GLOCs & MWPs & LLOCs & GLOCs & MWPs & LLOCs \\ \hline
				& 2559.29 & 1800.84 & 783.26 & 1873.58 & 1288.15 & 1310.83 & 1877.00 & 1364.69 & 1273.05  \\
				& 19027.58 & 3273.93 & 0.0 & 6822.54 & 3775.00 & 0.0 & 6824.22 & 3783.09 & 0.0 \\ 
				& 671163.75 & 0.0 & 395002.33 & 667209.59 & 0.0 & 459423.61 & 675095.75 & 0.0 & 425703.24  \\
				\hline
				& 18215.44 & 2318.44 & 65.43 & 6525.53 & 3422.37 & 28.88 & 6274.43 & 3037.82 & 53.48 \\
				& 2559.29 & 1800.84 & 783.26 & 1874.17 & 1282.66 & 1311.03 & 1877.50 & 1359.86 & 1273.47  \\
				& 2485.78 & 1797.36 & 537.03 & 2164.45 & 1723.96 & 852.58 & 2172.80 & 1721.73 & 818.55 \\ \hline
				& 18064.77 & 976.18 & 1039.22 & 4958.65 & 1007.88 & 1268.75 & 5055.28 & 1053.35 & 1222.04
		\end{tabular}}
		\caption{ARPA Grid Optimization Competition - GLOCs, MWPs, LLOCs [\$]}
		\label{tab:arpa_stats_ii}
	\end{table}
}{}

The results confirm the observations made for the other test systems. \footnote{Note that the ARPA-E test systems are much bigger in size, but only consider a single hour. For both the IEEE RTS system and the German bidding zone, we consider 24 consecutive hours. Therefore, the results reported in Table \ref{tab:arpa_stats_ii} has different orders of magnitude.} Min-MWP requires small make-whole payments, but incurs very large GLOCs and LLOCs. CH pricing minimizes GLOCs but incurs very large MWPs and LLOCs. Finally, IP pricing provides correct congestion signals, but requires large side-payments and penalties.

\end{document}